\documentclass[12pt]{iopart}

\usepackage{iopams,subeqn,setstack}  
\usepackage{bbm}
\newcommand{\eqref}[1]{\textup{(\ref{#1})}}
\newcommand{\dd}{\mathop{}\!\mathrm{d}}
\newcommand{\DD}{\mathrm{d}}
\newcommand{\Sl}{\mathop{\mathrm{sl}}\nolimits}
\newcommand{\ee}{\, \mathrm{e }}
\newcommand{\ch}{\mathop{\mathrm{ch}}\nolimits}
\newcommand{\sh}{\mathop{\mathrm{sh}}\nolimits}
\newcommand{\Th}{\mathop{\mathrm{th}}\nolimits}
\newcommand{\Sch}{\mathop{\mathrm{Sch}}\nolimits}

\makeatletter
\newcounter{parentequation}
\renewenvironment{subequations}{%
  \refstepcounter{equation}%
  \begingroup 
  \let\protect\@nx
  \edef\@tempa{\def\@nx\theparentequation{\theequation}}%
  \@xp\endgroup\@tempa
  \setcounter{parentequation}{\value{equation}}%
  \setcounter{equation}{0}%
  \def\theequation{\theparentequation$\alph{equation}$}%
  \ignorespaces
}{%
  \setcounter{equation}{\value{parentequation}}%
  \global\@ignoretrue
}
\makeatother
\renewcommand{\numparts}{\begin{subequations}}
\renewcommand{\endnumparts}{\end{subequations}}

\begin{document}

\title[The Schr\"odinger group and matrix orthogonal polynomials]{Representations of the Schr\"odinger group and matrix orthogonal polynomials}

\author{Luc Vinet$^1$, Alexei Zhedanov$^2$}
\address{$^1$Centre de recherches math\'ematiques, Universit\'e de Montr\'eal, C. P. 6128, succ.~Centre-ville, Montr\'eal, QC H3C 3J7, Canada}
\ead{luc.vinet@umontreal.ca}
\address{$^2$ Donetsk Institute for Physics and Technology, Donetsk 83114, Ukraine}
\ead{zhedanov@fti.dn.ua}

\begin{abstract}
The representations of the Schr\"odinger group in one space dimension are explicitly constructed in the basis of the harmonic oscillator states. These representations are seen to involve matrix orthogonal polynomials in a discrete variable that have Charlier and Meixner polynomials as building blocks. The underlying Lie-theoretic framework allows for a systematic derivation of the structural formulas (recurrence relations, difference equations, Rodrigues' formula etc.) that these matrix orthogonal polynomials satisfy.
\end{abstract}


\eqnobysec
 \maketitle

\section{Introduction}\label{vinet:introduction}
This paper purports to present an explicit construction of the representations of the 1-$D$ Schr\"odinger group, $\mathrm{Sch}_1$, in the basis of the state vectors of the harmonic oscillator. In physical terms, this entails for instance, the determination of transition amplitudes for oscillators subjected to external forces and varying frequencies. The ensuing results are of relevance to many areas of physics. Interestingly, our study also contributes to the mathematical theory of matrix orthogonal polynomials that is currently being actively developped. Indeed, as will be seen, the group-valued matrix elements of the $\mathrm{Sch}_1$--representation are expressed in terms of such matrix orthogonal polynomials (MOPs). Furthermore, the natural Lie--algebraic setting of our presentation allows to systematically obtain various structural formulas that characterize these MOPs. Much efforts are currently being deployed to obtain MOPs with such properties and their symmetry analysis has been rather limited. It is hence remarkable that our study provides in a most simple physical context, such a case-study of matrix orthogonal polynomials. Let us now give some background for our analysis.

\subsection{The Schr\"odinger group $\mathrm{Sch}_1$ and its algebra}\label{vinet:subsection1.1}

The Schr\"odinger group $\mathrm{Sch}_1$ is defined as the (maximal) group of symmetry transformations of the free one-dimensional Schr\"odinger equation:
\begin{eqnarray}\label{vinet:eq1.1} 
i \frac{\partial \psi}{\partial t} = -\frac12 \frac{\partial^2 \psi}{\partial x^2}.
\end{eqnarray}
This group was obtained by Niederer~\cite{vinet:ref1} in 1972, who found that the familiar Galilei transformations
 \begin{eqnarray}\label{vinet:eq1.2} 
(t,x) \to (t+b, x + vt + a)
\end{eqnarray}
are supplemented by
 \numparts
\begin{eqnarray}\label{vinet:eq1.3}
\mathrm{dilations} &: (t,x) \to (\mathrm{d}^2t, \dd x) \label{vinet:eq1.3a}
\\
\mathrm{and}&\nonumber\\
\mathrm{expansions} &: (t,x) \to \left( \frac{t}{1+\alpha t}, \frac{x}{1+\alpha t}\right). \label{vinet:eq1.3b}
\end{eqnarray}
\endnumparts
The infinitesimal symmetry generators $T$, $P$, $G$, $D$, $K$, and $E$ respectively associated to time and space translations, Galilei boosts, dilations, expansions and phase transformations are given by
\begin{eqnarray}\label{vinet:eq1.4} 
T &= i \partial_t, \qquad D = i\bigg(2t\partial_t + x\partial_x + \frac12\bigg),\nonumber \\
K &= -i\bigg(t^2 \partial_t + tx\partial_x + \frac{t}{2}\bigg) - \frac{x^2}{2}  \\[2\jot]
P &= -i\partial_x, \qquad G = -it\partial_x + x, \qquad E = 1.\nonumber
\end{eqnarray}
They are easily seen to form an algebra which is isomorphic to the semi-direct sum of $\Sl(2, \mathbb{R})$ and the Heisenberg algebra : $\Sl(2, \mathbb{R}) \ltimes H_1$. The generators $P$, $G$ and $E$ form a basis for the $H_1$ subalgebra, while $H$, $D$, and $K$ close onto $\Sl(2, \mathbb{R})$ under commutation.

\subsection{$\mathrm{Sch}_1$ and harmonic oscillators}\label{vinet:subsection1.2}

A year later, Niederer~\cite{vinet:ref2} again, studied the maximal symmetry group of the Schr\"odinger equation
\begin{eqnarray}\label{vinet:eq1.5} 
i \frac{\partial \psi}{\partial t} = -\frac{1}{2} \frac{\partial^2 \psi}{\partial x^2} + \frac{1}{2} x^2 \psi
\end{eqnarray}
with a harmonic potential to find that this equation admits an invariance group which  is isomorphic to $\mathrm{Sch}_1$.

Since this is the realization of $\mathrm{Sch}_1$ that we shall use in this paper, we shall record for reference, basic facts about quantum-mechanical harmonic oscillators.  As is well known, they are described in terms of the anhilation and creation operators
 \begin{eqnarray}\label{vinet:eq1.6} 
a = \frac{1}{\sqrt{2}} \left( x + \frac{\partial}{\partial x}\right), \qquad
a^+ = \frac{1}{\sqrt{2}} \left( x - \frac{\partial}{\partial x}\right)
\end{eqnarray}
that satisfy the Heisenberg algebra commutation relation
  \begin{eqnarray}\label{vinet:eq1.7} 
[a, a^+] = 1
\end{eqnarray}
and in terms of which
 \begin{eqnarray}\label{vinet:eq1.8}
 H =   - \frac{1}{2}  \frac{\partial^2}{\partial x^2} +   \frac{1}{2} x^2 = a^+ a + \frac{1}{2}.
 \end{eqnarray}
 We shall use throughout the standard representation that has for basis the normalized number vector states $\{ \mathopen\vert n  \mathclose\rangle, n=0,1,2,\ldots\}$, $\langle m\mathopen\vert n\mathclose\rangle = \delta_{mn}$, in which 
 \begin{eqnarray}\label{vinet:eq1.9}
a \mathopen\vert n \mathclose\rangle = \sqrt{n} \,   \mathopen\vert n-1 \mathclose\rangle,  \quad a^+ \mathopen\vert n \mathclose\rangle  = \sqrt{n+1}  \mathopen\vert n+1\mathclose\rangle, \  H  \mathopen\vert n \mathclose\rangle   = \left( n + \frac{1}{2}\right)  \mathopen\vert n \mathclose\rangle.
\end{eqnarray}
In the position representation, the wave functions are expressed in terms of the Hermite polynomials $H_n(x)$ (see~\ref{vinet:subsectionA.1}):
 \begin{eqnarray}\label{vinet:eq1.10}
\Phi_n(x) = \langle x \mathopen\vert n \mathclose\rangle = \frac{1}{\pi^{1/4}2^{n/2}\sqrt{n!}}\, \ee^{-x^2/2} H_n(x).
\end{eqnarray}
We shall also make use of the coherent states $\mathopen| z \mathclose\rangle$, $z \in \mathbb{C}$ defined as the eigenstates of the annihilation operator $a$. These have the following properties:
 \numparts
 \begin{eqnarray}\label{vinet:eq1.11}
 \mathopen| z \mathclose\rangle = \ee^{za^+}  \mathopen| 0 \mathclose\rangle, \qquad  \mathopen| z \mathclose\rangle  = \sum_{n=0}^{\infty} \frac{z^n}{\sqrt{n!}}  \mathopen| n \mathclose\rangle , \qquad \ee^{\beta a^+} \mathopen| z \mathclose\rangle  = \mathopen| z + \beta \mathclose\rangle, \label{vinet:eq1.11a}\\[2\jot]
  a\mathopen| z \mathclose\rangle  = z \mathopen| z \mathclose\rangle, \qquad f(a) \mathopen| z \mathclose\rangle  = f(z) \mathopen| z \mathclose\rangle, \label{vinet:eq1.11b}\\[2\jot]
  \mathopen\langle n \mathclose| z\rangle =  \frac{z^n}{\sqrt{n!}}, \qquad \langle y \mathopen| z\mathclose{\rangle} =\ee^{\bar{y} z}
\label{vinet:eq1.11c}
\end{eqnarray}
\endnumparts
where $f$ is an arbitrary (smooth) function in (\ref{vinet:eq1.11b}).

Let us now give an indication as to why the invariance group of (\ref{vinet:eq1.5}) is isomorphic to $\mathrm{Sch}_1$. The symmetry transformations of space and time are obviously different in this case than those that map solutions of the free equation onto themselves, it is however easy to see that their generators span the same abstract algebra, that is $\Sl(2, \mathbb{R}) \ltimes H_1$.

We already noted that $a$, $a^+$ and $I$ form a basis for the Heisenberg algebra $H_1$. Similarly the bilinears $a^2$, $a^+a$ and $(a^+)^2$ realize the $\Sl(2,\mathbb{R})$ commutation rules. Observe that the sets of state $\{ |2n\rangle, n=0,1,\ldots\}$ and $\{|2n+1\rangle, n=0, 1,\ldots\}$ with even and odd quanta, will separately transform onto themselves under the action of those bilinears; as a matter of fact, they span two different irreducible discrete representation series of $\Sl(2, \mathbb{R})$. It is straightforward to check that together the linears and bilinears in $a$ and $a^+$ realize the semi-direct sum $\Sl(2,\mathbb{R})\ltimes H_1$.

Consider the symmetry generators (\ref{vinet:eq1.4}) of the free Schr\"odinger equation at $t=0$
\begin{eqnarray} \label{vinet:eq1.12}
\eqalign{P(0) = -i\partial_x \qquad G(0) = x, \qquad E(0) = I,  \cr
T(0) = -\frac{1}{2} \partial_x^2, \qquad D(0) =  x \partial_x + \frac{1}{2}, \qquad K(0) = -\frac{x^2}{2}.}
\end{eqnarray}
We know they form a $\Sl(2,\mathbb{R}) \ltimes H_1$ algebra. They also clearly span the set of linears and bilinears in $a$ and $a^+$ at since
\begin{eqnarray} 
P(0) = -\frac{i}{\sqrt{2}}(a-a^+), \qquad
G(0) = \frac{1}{\sqrt{2}} (a+a^+), \qquad
E(0) = aa^+ - a^+ a,\nonumber\\
T(0) = - \frac{1}{4}(a^2 + (a^+)^2 - aa^+ - a^+a), \qquad
D(0) = \frac{1}{2} (a^2 - (a^+)^2), \label{vinet:eq1.13}\\
K(0) = -\frac14 (a^2 + (a^+)^2 + aa^+ + a^+a). \nonumber
\end{eqnarray}
If we take these operators generically denoted by $X(0)$ and evolve them in time using the harmonic oscillator Hamiltonian $H$ given in (\ref{vinet:eq1.8}) (rather than the free Hamiltonian) according to
\begin{eqnarray} \label{vinet:eq1.14}
X(t) = \ee^{iHt} X(0) \ee^{-iHt};
\end{eqnarray}
this will obviously yield constants of motion,
\begin{eqnarray} \label{vinet:eq1.15}
\frac{\dd X}{\dd t} (t) = \frac{\partial X}{\partial t}(t) - i [H, X(t) ] = 0,
\end{eqnarray}
that is, symmetry generators. Since the time evolution (\ref{vinet:eq1.14}) is given by a unitary automorphism, the time-dependent symmetry generators will satisfy at arbitrary $t$ the same commutation relations than at $t=0$, namely those of $\Sl(2,\mathbb{R}) \ltimes H_1$. The symmetry transformation of $(t,x)$ that leave (\ref{vinet:eq1.15}) invariant can then be obtained by exponentiating these generators. This explains why the symmetry groups of the Schr\"odinger equation for the free particle and for the harmonic oscillator are structurally identical.

\subsection{Representations of the $\Sch_1$ on oscillator states: the central object}\label{vinet:subsection1.3}

The projective representations of the Schr\"odinger group in $n$ dimensions $\Sch_n$ have been classified by Perroud~\cite{vinet:ref3}. More recently, an infinite-dimensional extension of the Schr\"odinger algebra known as the Schr\"odinger--Virasoro algebra has been constructed and studied~\cite{vinet:ref3,vinet:ref4,vinet:ref5}. This extended algebra has been used in particular, to analyze anisotropic critical systems and to formulate a non-relativistic version of the AdS/CFT correspondance.  The book of Roger and Unterberger~\cite{vinet:ref6} covers the mathematics of these developments; moreover, a short review of the many physical situations (old and new) exhibiting Schr\"odinger symmetry, with references to the literature, is provided in the introduction. 

In the following, we shall be interested in the explicit determination of the matrix elements for the unitary representation of $\Sch_1$ in the basis $\{|n\rangle, n = 0,1,2,\ldots\}$ of the oscillator quantum states. A priori, 6 parameters are needed to specify a general $\Sch_1$ group element. However, since the factor $\ee^{i\in H} \cdot \ee^{i\alpha E}$ is diagonal and hence trivial in the oscillator basis, it will not be included. With this understanding, we shall use the following parametrization for a general group element of $\Sch_1$:
\begin{eqnarray} \label{vinet:eq1.16}
 S(v,w) = \ee^{(va - \bar{v} a^+)} \ee^{(wa^2 - \overline{w} (a^+)^2)/2} \in \Sch_1
 \end{eqnarray}
where $v$ and $w$ are complex numbers (equivalent to four real parameters) and $\bar{v}$ and $\overline{w}$ their complex conjugate. Note that $S$ is a unitary operator : $S^+ = S^{-1}$.

Our central objects of study will hence be the matrix elements
\begin{eqnarray} \label{vinet:eq1.17}
\, \psi_{n,k} = \langle k |S|n\rangle = \langle k|\ee^{(va-\bar{v}a^+)} \ee^{(wa^2 - \overline{w}(a^+)^2)/2}|n\rangle.
 \end{eqnarray}
 
 As already mentioned, we shall show that these quantities are naturally expressed in terms of $2 \times 2$ matrix orthogonal polynomials $\mathcal{P}_n(k)$ of the discrete variable $k$. The reader will also have realized the relations with the theory of coherent and squeezed states \cite{vinet:ref7} that our analysis will entail.
 
 \subsection{Matrix orthogonal polynomials}\label{vinet:subsection1.4}
 
 Introduced by Krein some 60 years ago, the theory of matrix orthogonal polynomials (MOPs) had not much progressed until a few years ago when Duran, Gr\"unbaum and coworkers gave it a significant impetus with the discovery of families of MOPs that satisfy second-order differential equations. A survey of these advances can be found in \cite{vinet:ref8} and our study adds to this corpus of results. Let us also signal Ref.~\cite{vinet:ref9a}, were the relation between polynomials obeying higher-order recurrence relations and matrix orthogonal polynomials was identified.
 
 We offer below basic definitions and properties of matrix orthogonal polynomials. (The reader is referred to \cite{vinet:ref9} for a broad exposition.) Let $(\mathcal{P}_n)_n$ be a sequence of polynomials with $N\times N$ matrix coefficients (or equivalently, a sequence of $N\times N$ matrices with polynomials of degree $n$ or less as entries). In the case where these $\mathcal{P}_n$ depend on a continuous variable $t$, they will be orthogonal if there exists a suitably defined weight matrix $W(t)$ such that
 \begin{eqnarray} \label{vinet:eq1.18}
\int \mathcal{P}_n(t) \dd W(t) \mathcal{P}_m^+ (t) = h_n \delta_{n,m}(I)_{N\times N}.
 \end{eqnarray}
Like in the scalar case, a sequence of matrix orthogonal polynomials satisfy a 3-term recurrence relation:
\begin{eqnarray} \label{vinet:eq1.19}
t\mathcal{P}_n (t) = A_{n+1}\mathcal{P}_{n+1}(t) + B_n(t)\mathcal{P}_n(t) + A_n^+\mathcal{P}_{n-1}(t), \qquad n \ge 0
\end{eqnarray}
 where $\mathcal{P}_{-1}(t) = 0$, $A_n$ are non-singular matrices and $B_n$ are Hermitian $B_n^+ = B_n$. Strikingly, as reviewed in \cite{vinet:ref8}, for some families of MOPs, the $\mathcal{P}_n$ have been shown to be eigenvectors of right second-order differential operators, that is to satisfy and equation of the form
 \begin{eqnarray} \label{vinet:eq1.20}
 \frac{\DD^2}{\dd t^2} \mathcal{P}_n(t) A_2(t) + \frac{\dd}{\dd t} \mathcal{P}_n(t) A_1(t) + \mathcal{P}_n(t) A_0(t) = \Gamma_n\mathcal{P}_n(t)
  \end{eqnarray}
 where the differential coefficients $A_2$, $A_1$ and $A_0$ are matrix polynomials independent of $n$, of degree not higher than 2, 1 and 0 respectively and $\Gamma_n$ are Hermitian matrices. When the MOPs $\mathcal{P}_n$ are functions of a discrete variable, as will be the case in our example, a discrete sum replaces the integral in the orthogonality relations (\ref{vinet:eq1.18}) and one would wish to identify families of matrix polynomials that are eigenvectors of right-difference operators. The reader is referred to \cite{vinet:ref10} for some specifics of matrix orthogonal polynomials of a discrete variable.
 
 \subsection{Outline}\label{vinet:subsection1.5}
 
 The paper is organized as follows. The recurrence relation that the matrix elements $\psi_{n,k}$ satisfy is obtained in Section~\ref{vinet:section2}. Upon analysis, it will be seen to imply that the $\psi_{n,k}$ are given in terms of $2 \times 2$ matrix orthogonal polynomials in $k$, $\mathcal{P}_n(k)$. The weight matrix $W(k)$ for these MOPs is derived in Section~\ref{vinet:section3} from the unitarity of the representation and expressed using $\psi_{0,k}$ and $\psi_{1,k}$ solely. Special attention is given to those particular matrix elements in  Section A. On the one hand, Hermite polynomials will be seen to enter in their explicit formula and on the other hand, they will be shown to form a $k$-indexed doublet for which raising and lowering matrices will be constructed. This will provide the necessary tools to describe the bispectrality of the $\mathcal{P}_n(k)$ in Section~\ref{vinet:section5} and to obtain the difference equation that they obey. Section~\ref{vinet:section6} will be dedicated to the ladder relations and the Rodrigues'  formula. It will be further shown in Section~\ref{vinet:section7} that the matrix elements $\psi_{n,k}$ can be written as a ``discrete'' convolution involving Charlier and Meixner polynomials. Generating functions will be the object of Section~\ref{vinet:section8}. Introducing the 2-variable generating function of Perelomov and Popov~\cite{vinet:ref11} will bring to light the close connection between the matrix elements $\psi_{n,k}$ and Hermite polynomials in two variables and will provide as a corollary intricate relations between the orthogonal polynomials involved. Finally the transformation law of the standard Hermite polynomials under an affine change of the independent variable is obtained as a by-product in Section~\ref{vinet:section9} upon passing to the position representation. We close with concluding remarks and gather in an appendix the properties of orthogonal polynomials that are used in the text.
 
\section{Recurrence relations} \label{vinet:section2}

We shall initiate our analysis of the representation of $\Sch_1$ in the oscillator basis by determining the recurrence relation that the matrix elements $\psi_{n,k}$ of (\ref{vinet:eq1.17}) satisfy. We shall follow an approach already used in \cite{vinet:ref12}.

We obviously have
\begin{eqnarray} \label{vinet:eq2.1}
\langle k \vert a^+a S \vert n \rangle = k \langle k \vert S \vert n \rangle = k \, \psi_{n,k}
  \end{eqnarray}
  and also
\begin{eqnarray} \label{vinet:eq2.2}
  \langle k \vert a^+a S \vert n \rangle = \langle k \vert SS^{-1} a^+aS\vert n\rangle.
    \end{eqnarray}
 With $S$ given by (\ref{vinet:eq1.16}), using the Baker--Campbell--Haussdorf formula
  \begin{eqnarray} \label{vinet:eq2.3}
  \ee^X Y \ee^{-X} = Y + [X,Y] + \frac{1}{2!}\big[X[X,Y]\big] + \frac{1}{3!}\big[X\big[X[X,Y]\big]\big] + \ldots
  \end{eqnarray}
  we can straightforwardly obtain the automorphism of $a^+a$ (a $\Sch_1$ algebra element) under $S$. In the following we shall use the four real parameters $\rho$, $\theta$, $\sigma$, $\delta$ related to $v$ and $w$ as follows;
\begin{eqnarray} \label{vinet:eq2.4}
 v = \sigma \ee^{i\delta}, \qquad w = \rho \ee^{i\theta}.
 \end{eqnarray}
We hence find
 \numparts
\begin{eqnarray}\label{vinet:eq2.5}
S^{-1}aS &= a \ch \rho - a^+ \ee^{-i\theta} \sh \rho - \sigma \ee^{-i\delta}, \label{vinet:eq2.5a}
\\
 S^{-1}a^+S &= a^+\ch \rho - a \ee^{i\theta} \sh \rho - \sigma \ee^{i\delta}; \label{vinet:eq2.5b}
\end{eqnarray}
\endnumparts
from where it follows that
\begin{eqnarray} \label{vinet:eq2.6}      
\fl S^{-1} a^+aS   =  a^+a \ch 2\rho + \frac12 \ch 2\rho + \sigma^2 - \frac12 
   -\frac12 a^2 \sh 2\, \rho \ee^{i\theta} - \frac12 (a^+)^2 \sh 2\, \rho \ee^{-i\theta}\nonumber \\
    + a \sigma (\ee^{i(\theta-\delta)}\sh\rho - \ee^{i\delta}\ch \rho)  
  + a^+\sigma(\ee^{-i(\theta - \delta)} \sh \rho - e^{-i\delta} \ch \rho).
\end{eqnarray}
Using (\ref{vinet:eq2.6}) in (\ref{vinet:eq2.2}) and the actions of $a$ and $a^+$ on $\vert n \rangle$ given in (\ref{vinet:eq1.9}), one finds upon combining (\ref{vinet:eq2.2}) and (\ref{vinet:eq2.1}) the following 5-term recurrence relation for the matrix elements $\psi_{n,k}$:
\begin{eqnarray}
\fl k\, \psi_{n,k}  = \bigg[ \bigg(n + \frac12\bigg)\ch 2\rho + \sigma^2 - \frac12\bigg] \, \psi_{n,k}\nonumber\\
 -\frac12 \sqrt{n(n-1)} \sh 2\, \rho \ee^{i\theta} \psi_{n-2,k} - \frac12\sqrt{(n+1)(n+2)} \,  \sh 2\, \rho \ee^{i\theta} \psi_{n+2, k}\nonumber\\  
 + \sqrt{n} \,  \sigma (\ee^{i(\theta-\delta)}\sh \rho - \ee^{i\delta}\ch \rho) \, \psi_{n-1,k}\nonumber\\
   + \sqrt{{n+1}} \sigma (\ee^{-i(\theta-\delta)}\sh \rho - \ee^{-i\delta}\ch \rho) \, \psi_{n+1,k}. \label{vinet:eq2.7}    
  \end{eqnarray}
The analysis of this recurrence relation performed by setting $n=0,1,2,\ldots$ and so on, shows that $\psi_{2n,k}$ and $\psi_{2n+1,k}$ are separately given in terms of $\psi_{0,k}$ and $\psi_{1,k}$ through expressions involving four sets of polynomials $P_n(k)$, $Q_n(k)$, $\widetilde{P}_n(k)$ and $\widetilde{Q}_n(k)$ of degree $n$ in the variable $k$:
 \numparts 
\begin{eqnarray} \label{vinet:eq2.8} 
\psi_{2n,k} &= P_n(k) \, \psi_{0,k} + Q_{n-1}\psi_{1,k} \label{vinet:eq2.8a} \\
\psi_{2n+1,k} &= \widetilde{P}_n(k) \, \psi_{0,k} + \widetilde{Q}_{n}\psi_{1,k}. \label{vinet:eq2.8b}
\end{eqnarray}
\endnumparts
Note that it is a polynomial of degree $n-1$ that is the factor of $\psi_{1,k}$ in (\ref{vinet:eq2.8a}). The initial conditions are
 \numparts
\begin{eqnarray}\label{vinet:eq2.9}
P_0(k) &=   \widetilde{Q}_{0}(k) = 1 \label{vinet:eq2.9a} \\
Q_{-1}(k) &= \widetilde{P}_0 (k)=0. \label{vinet:eq2.9b}
\end{eqnarray}
\endnumparts
 Observe also that in the two special cases $\rho = 0$ and $\sigma = 0$, the recurrence relation (\ref{vinet:eq2.7}) reduces to 3-term recurrence relations. In the latter case $(\rho = 0)$, (\ref{vinet:eq2.7}) implies in particular that $\psi_{1,k}$ is given in terms of $\psi_{0,k} \colon \psi_{1,k} =(\sigma - k/\sigma)\ee^{i\delta} \psi_{1,k}$. These two special cases will be treated in Section~\ref{vinet:section7}. Until then and unless specified, we take $\rho \ne 0 \ne \sigma$. When this is so, the two initial ``values'' $\psi_{0,k}$ and $\psi_{1,k}$ of the recurrence solutions are independent.
 
 In view of   (\ref{vinet:eq2.8a}, \ref{vinet:eq2.8b}), it is natural to unify $\psi_{2n,k}$ and $\psi_{2n+1,k}$ in a 2-vector that we shall denote $\Psi_{n,k}$:
\begin{eqnarray} \label{vinet:eq2.10} 
\, \Psi_{n,k} = 
\left(
\begin{array}{l}
\psi_{2n,k}\\
\psi_{2n+1,k}
\end{array}
\right).
\end{eqnarray} 
  
Let us now define the quantities $\xi_n$, $\eta_n$ and $\zeta_n$ by
 \numparts
\begin{eqnarray}\label{vinet:eq2.11}
\xi_n &= -\frac12 \sqrt{(n-1)n} \sh 2\, \rho \ee^{-i\theta} \label{vinet:eq2.11a} \\
\eta_n &= \sqrt{n} \,  \sigma (\ee^{i(\delta -\theta)}\sh \rho - \ee^{-i\delta} \ch \rho)  \label{vinet:eq2.11b} \\
\zeta_n &= \bigg( n + \frac12 \bigg)\ch 2\rho + \sigma^2 - \frac12 \label{vinet:eq2.11c} 
\end{eqnarray}
\endnumparts
    and introduce the $2\times 2$ matrices $A_n$ and $B_n$:
 \numparts
\begin{eqnarray}\label{vinet:eq2.12}
A_n &=\left(
 \begin{array}{c c}
  \xi_{2n} &0\\
  \eta_{2n} &\xi_{2n+1}
  \end{array} \right)
  \label{vinet:eq2.12a}
 \\[2\jot]
B_n &=
 \left(\begin{array}{l l}
  \xi_{2n} &\eta_{2n+1}\\
  \overline{\eta}_{2n+1} &\zeta_{2n+1}
  \end{array}
  \right).\label{vinet:eq2.12b}
    \end{eqnarray}
\endnumparts
 Note that $B_n$ is Hermitian $B_n^+ =B_n$.
 
  It is then straightforward to see that the 5-term scalar recurrence relation  (\ref{vinet:eq2.7}) can be written as the following 3-term recurrence relation for the 2-vector $\Psi_{n,k}$:
   \begin{eqnarray} \label{vinet:eq2.13}
   k \, \Psi_{n,k} = A_{n+1} \Psi_{n+1,k} + B_n \, \Psi_{n,k} + A_n^+  \Psi_{n-1,k}.
  \end{eqnarray}
  If we bring in the matrix polynomial
\begin{eqnarray}
\mathcal{P}_n &=\left(
 \begin{array}{l l}  
  P_n(k) &Q_{n-1}(k)\\[2\jot]
  \widetilde{P}_n(k) &\widetilde{Q}_n(k)
   \end{array}
  \right)\label{vinet:eq2.14}
    \end{eqnarray}
it is also immediate to observe that the recurrence solution (\ref{vinet:eq2.8a}, \ref{vinet:eq2.8b})--(\ref{vinet:eq2.9a}, \ref{vinet:eq2.9b}) now takes the following form
  \begin{eqnarray} \label{vinet:eq2.15}
  \, \Psi_{n,k} = \mathcal{P}_n(k) \Psi_{0,k}
  \end{eqnarray}
  with initial condition
   \begin{eqnarray} \label{vinet:eq2.16}
   \mathcal{P}_0(k)= {\mathbbm{1}}_{2 \times 2}.
  \end{eqnarray} 
  
  Evidently,  (\ref{vinet:eq2.15}) and (\ref{vinet:eq2.16}) could have been obtained directly as the recurrence solution of  (\ref{vinet:eq2.13}). Now remembering that for generic $\rho$ and $\sigma$, $\Psi_{0,k}$ has no annihilator (since otherwise $\psi_{0,k}$ and $\psi_{1,k}$ would not be independent), using  (\ref{vinet:eq2.15}) in  (\ref{vinet:eq2.13}) one arrives at the conclusion that the matrix polynomials $\mathcal{P}_n(k)$ obey themselves a 3-term recurrence relation, that is
\begin{eqnarray} \label{vinet:eq2.17}
k\mathcal{P}_n(k) = A_{n+1}\mathcal{P}_{n+1}(k) + B_n \mathcal{P}_{n,k} (k) + A_n^+ \mathcal{P}_{n-1}(k), \qquad \mathcal{P}_0(k) = \mathbbm{1}.
  \end{eqnarray}  
 From the short introduction given in Subsection~\ref{vinet:subsection1.4}, one thus concludes that the polynomials $\mathcal{P}_n(k)$ are orthogonal. This is the central result of the paper, namely that the matrix elements (\ref{vinet:eq1.17}) of the $\Sch_1$ representation are given in terms of the matrix orthogonal polynomials $\mathcal{P}_n(k)$ that are of degree $n$ in the discrete variable $k$. Much of the following will be aimed at characterizing these $\mathcal{P}_n(k)$.
  
 \section{Weight matrix}\label{vinet:section3}
 
 We shall determine in this section the $2 \times 2$ weight matrix $W(k)$ for which the polynomials $\mathcal{P}_n(k)$ defined by the recurrence relation (\ref{vinet:eq2.17}) are orthogonal. We shall see that it is readily provided by the unitarity of the representation.
 
 Since $S^+ = S^{-1}$, we have
   \begin{eqnarray} \label{vinet:eq3.1}
   \delta_{n,m} &= \langle m \vert S^+ S\vert n \rangle\\
   &= \sum_{k=0}^{\infty} \langle m \vert S^+\vert k\rangle \, \langle k \vert S \vert n \rangle = \sum_{k=0}^{\infty} \, \psi_{n,k} \overline{\psi}_{m,k}.
  \end{eqnarray}
  If we use the 2-vector $\Psi_{n,k}$, this identity can be cast in the form
    \begin{eqnarray} \label{vinet:eq3.2}
\sum_{k=0}^{\infty} \, \Psi_{n,k} \Psi^+_{m,k} = \delta_{n,m} {\mathbbm{1}}_{2\times 2}.
  \end{eqnarray}
  Using the expression (\ref{vinet:eq2.15}) giving $\Psi_{n,k}$ in terms of $\mathcal{P}_n(k)$, we see that (\ref{vinet:eq3.2}) implies
  \begin{eqnarray} \label{vinet:eq3.3}
  \sum_{k=0}^{\infty} \mathcal{P}_n(k) \Psi_{0,k} \Psi_{0,k}^+ \mathcal{P}_m(k)^+ = \delta_{n,m} {\mathbbm{1}}_{2\times 2}.
  \end{eqnarray}  
  This is the orthogonality relation we were looking for (see (\ref{vinet:eq1.18}) in the continuous case), since it can be written
    \begin{eqnarray} \label{vinet:eq3.4}
    \sum_{k=0}^{\infty} \mathcal{P}_n(k) W(k) \mathcal{P}_m^+(k) = \delta_{n,m} {\mathbbm{1}}_{2\times 2},
  \end{eqnarray}
  where $W(k)$ the $2 \times 2$ weight matrix is given by
    \begin{eqnarray} \label{vinet:eq3.5}
    W(k) = \Psi_{0,k} \Psi_{0,k}^+ =
    \left(\begin{array}{l l}
 |\psi_{0,k}|^2  &\psi_{0,k} \overline{\psi_{1,k}}\\[3\jot]
  \overline{\psi_{0,k}}\psi_{1,k}  & 
  |\psi_{1,k}|^2  
  \end{array}
  \right). 
  \end{eqnarray}
  In the next section, we shall focus on the matrix elements $\psi_{0,k}$ and $\psi_{1,k}$, the initial values of the recurrence relations, that form as we see the elements of the weight matrix.

 \section{\mathversion{bold}The matrix elements $\psi_{0,k}$ and $\psi_{1,k}$} \label{vinet:section4}
    
    We wish to have explicit expressions for $\psi_{0,k}$ and $\psi_{1,k}$. To that end, we shall first compute the vacuum or ground state expectation values $\psi_{0,1} = \langle 0| S |0\rangle$.
    
    \subsection{$\psi_{0,0}$}\label{vinet:subsection4.1}
    In order to carry out the calculation of this quantity we need so-called ``disentangling theorems'', that is tools to factor the group element $S$ as ordered products of exponentials solely of monomials in the creation and annihilation operators. For the first factor of $S$, this is easily achieved using the formula
   \begin{eqnarray} \label{vinet:eq4.1}
   \ee^{A+B} = \ee^{-[A,B]/2}\ee^A \cdot \ee^B
  \end{eqnarray}
  known to be true when $A$ and $B$ commute with their commutator. One has thus
    \begin{eqnarray} \label{vinet:eq4.2}
    \ee^{(va-\bar{v}a^+)} = \ee^{-v\bar{v}/2}\ee^{-\bar{v}a^+}\ee^{va}.
  \end{eqnarray}
  
  The factorization of the second factor in $S$ is more involved but fairly well known in view of its wide use. It is performed in details in ref.~\cite{vinet:ref13}. In a nutshell, the approach therein is as follows. One lets
\begin{subequations}  
\label{vinet:eq4.3}
    \begin{eqnarray} \label{vinet:eq4.3a}
    S_1(\lambda) = \ee^{\lambda(wa^2 - \overline{w} (a^+)^2)}
   \end{eqnarray}
  and
      \begin{eqnarray} \label{vinet:eq4.3b}
      S_2(\lambda) = \ee^{f(\lambda)(a^+)^2} \ee^{q(\lambda)(a^+a +1/2)} \ee^{h(\lambda)a^2}
   \end{eqnarray}
   \end{subequations}
    and postulate that $S_1(\lambda) = S_2(\lambda)$. Differentiating both sides with respect to the parameter $\lambda$, using the  B-C-H  formula (\ref{vinet:eq2.3}) and the posited identity, yields differential equations for the functions $f(\lambda)$, $g(\lambda)$ and $h(\lambda)$. Solving with the appropriate initial conditions gives the following identity when $\lambda = 1$:
 \begin{eqnarray} \label{vinet:eq4.4}
\fl  \ee^{[wa^2 - \overline{w} (a^+)^2]/2}  = 
   \exp\bigg[ -\frac12 \ee^{-i\theta} \Th \rho\,  (a^+)^2\bigg]\nonumber\\[2\jot]
 \cdot   \exp \bigg[ -\ln (\ch \rho) \bigg(a^+a +\frac12\bigg)\bigg]\nonumber\\[2\jot]
 \cdot  \exp \bigg[ \frac12 \ee^{i\theta} \Th \rho\,  a^2 \bigg], \qquad (w = \rho \ee^{i\theta}).
  \end{eqnarray}
  We are now equipped to compute $\psi_{0,0}$. Using (\ref{vinet:eq4.2}) and (\ref{vinet:eq4.4}), we readily see that
  \begin{eqnarray}
  \psi_{0,0}
  &= \langle 0| S|0\rangle\nonumber \\
  &= \ee^{-\sigma^2/2} (\ch \rho)^{-1/2} \langle 0 | \exp(\sigma \ee^{i\delta}a)\exp \bigg[-\frac12 (\ee^{-i\theta} \Th \rho) (a^+)^2\bigg] |0\rangle\nonumber\\
  &= \ee^{-\sigma^2/2}(\ch \rho)^{-1/2} \exp \bigg( -\frac12 \sigma^2 \ee^{i(2\delta - \theta)} \Th \rho \bigg).  \label{vinet:eq4.5}
  \end{eqnarray}  
  
  \subsection{$\psi_{0,k}$} \label{vinet:subsection4.2}
  
  A recurrence relation for the matrix elements $\psi_{0,k}$ is easily obtained from observing that
  \begin{eqnarray} \label{vinet:eq4.6}
\langle k |S a |0\rangle = \langle k |S aS^{-1}S |0\rangle = 0.
  \end{eqnarray}  
 In the usual fashion one finds that
   \begin{eqnarray} \label{vinet:eq4.7}
   SaS^{-1} = \ch \rho\, a + \e^{-i\theta} \sh \rho\, a^+ + \sigma(\ee^{-i\delta}\ch \rho + \ee^{i(\delta-\theta)} \sh \rho),
  \end{eqnarray} 
  which leads to
    \begin{eqnarray} \label{vinet:eq4.8}
  \fl  \sqrt{k+1}\,  \ch \rho\, \psi_{0,k+1}  + \e^{-i\theta} \sh \rho \sqrt{k}\,    \psi_{0,k-1}\nonumber \\
     + \sigma (\ee^{-i\delta} \ch \rho + \ee^{i(\delta-\theta)}\sh \rho) \, \psi_{0,k} = 0
  \end{eqnarray}
      when employed in (\ref{vinet:eq4.6}). This 3-term recurrence relation has a simple solution. Let
  \begin{eqnarray} \label{vinet:eq4.9}
  \psi_{0,k} = \frac{1}{\sqrt{k!}} \bigg( \frac12 \ee^{-i\theta} \Th \rho\bigg)^{k/2} p_k\, \psi_{0,0}
  \end{eqnarray}     
  with $p_0 =1$. Upon substituting in (\ref{vinet:eq4.8}) one finds
  \begin{eqnarray} \label{vinet:eq4.9'}
  p_{k+1} + 2k p_{k-1} - 2sp_k = 0
  \end{eqnarray}    
    with
      \begin{eqnarray} \label{vinet:eq4.10}
      s = - \frac{\sigma}{\sqrt{2}}  [ \ee^{i(\delta-\theta/2)} \Th^{1/2} \rho + \ee^{-i(\delta - \theta/2)} \Th^{-1/2} \rho].
  \end{eqnarray}
 Comparing with the 3-term recurrence relation  (\ref{vinet:eqA.1}) of the Hermite polynomials, we see that $p_k = H_k(y)$ and hence that
  \begin{eqnarray} \label{vinet:eq4.11}
  \psi_{0,k} = \frac{1}{\sqrt{k!}} \bigg( \frac12  \ee^{-i\theta} \Th \rho\bigg)^{k/2} H_k(s) \, \psi_{0,0}
  \end{eqnarray} 
    where $H_k(s)$ are Hermite polynomials of the variable $s$ given in  (\ref{vinet:eq4.10}) and $\psi_{0,0}$ is as in (\ref{vinet:eq4.5}).
   \subsection{$\psi_{1,k}$}\label{vinet:subsection4.3}
    
    The matrix elements $\psi_{1,k}$ can in turn be obtained from the relation
  \begin{eqnarray} \label{vinet:eq4.12}
 \langle k | aS|0\rangle = \sqrt{k+1}\,  \langle k+1 |S|0 \rangle = \langle k|SS^{-1} aS|0\rangle.
  \end{eqnarray}    
    Using (\ref{vinet:eq2.5a}), it implies
  \begin{eqnarray} \label{vinet:eq4.13}
  \sqrt{k+1}\,  \psi_{0,k+1} = -\ee^{-i\theta}\sh \rho\, \psi_{1,k} - \sigma \ee^{-i\delta} \psi_{0,k}
  \end{eqnarray}    
in other words, that $\psi_{1,k}$ can be expressed in terms of $\psi_{0,k}$ and $\psi_{0,k+1}$:
    \begin{eqnarray} \label{vinet:eq4.14}
\psi_{1,k} = -\frac{\ee^{i\theta}}{\sh \rho} [\sigma \ee^{-i\delta} \psi_{0,k} + \sqrt{k+1}\,  \psi_{0,k+1}].
  \end{eqnarray}     
 Recalling expression (\ref{vinet:eq4.11}) for $\psi_{0,k}$, one has for $\psi_{1,k}$:
  \begin{eqnarray}
 \fl  \psi_{1,k} 
    = -\frac{1}{\sqrt{k!}} \cdot \frac{1}{\sh \rho} \bigg( \frac{\ee^{-i\theta}}{2} \Th \rho\bigg)^{k/2}\nonumber\\
    \times \bigg\{ \sigma \ee^{i(\theta-\delta)} H_k (s) +  \bigg( \frac12 \ee^{i\theta} \Th \rho\bigg)^{1/2} H_{k+1}(s) \bigg\} \cdot \psi_{0,0}. \label{vinet:eq4.15}
  \end{eqnarray}
Let us point out that this result could have been obtained using a different route. One could have started from    
  \begin{eqnarray} \label{vinet:eq4.16}
  \psi_{1,k} = \langle k|Sa^+|0\rangle = \langle k |Sa^+S^{-1} S|0\rangle.
  \end{eqnarray}    
  Using the Hermitian conjugate of (\ref{vinet:eq4.7}), this is seen to entail
 \begin{eqnarray} \label{vinet:eq4.17}
 \fl \psi_{1,k}  = \ch \rho \sqrt{k}\, \,  \psi_{0,k-1} + \ee^{i\theta}\sh \rho \sqrt{k+1}\,  \psi_{0,k+1}\nonumber\\
   + (\sigma \ee^{i\delta} \ch \rho +\sigma \ee^{-i(\delta-\theta)} \sh \rho) \, \psi_{0,k}.
  \end{eqnarray}  
  The 3-term recurrence relation (\ref{vinet:eq4.8}) for $\psi_{0,k}$ can then be called upon to eliminate $\psi_{0,k-1}$ in (\ref{vinet:eq4.17}) and recover (\ref{vinet:eq4.14}).
  
  The upshot of Subsections~\ref{vinet:subsection4.2} and \ref{vinet:subsection4.3} is that $\psi_{0,k}$ and $\psi_{1,k}$ and hence the weight matrix $W(k)$     (\ref{vinet:eq3.5}) of the matrix polynomials $\mathcal{P}_n(k)$ are given in terms of Hermite polynomials in the variable $s$ (a function of the group parameters). It is also manifest from(\ref{vinet:eq4.11}) and (\ref{vinet:eq4.15}) that $\psi_{0,k}$ and $\psi_{1,k}$ are independent when $\rho \ne 0 \ne \sigma$.
  
\subsection{Raising and lowering matrices}\label{vinet:subsection4.4}

For purposes that will be clear in the next two sections, it is relevant to have explicit ladder relations for $\psi_{0,k}$ and $\psi_{1,k}$. As it turns out, the relations ensuing from the identities
 \numparts\label{vinet:eq4.18}
  \begin{eqnarray} 
 \sqrt{k+1}\,  \psi_{i,k+1} &= \langle k |aS|i\rangle = \langle k |SS^{-1} aS|i\rangle, \label{vinet:eq4.18a}\\
 \sqrt{k}\,  \psi_{i,k-1} &= \langle k |a^+S|i\rangle = \langle k |SS^{-1} a^+S|i\rangle, \qquad i = 0,1 \label{vinet:eq4.18b}
  \end{eqnarray}
\endnumparts
  allow to express $\psi_{i, k\pm 1}$ in terms of $\psi_{0,k}$ and $\psi_{1,k}$. Let us see how this goes.
  
  Equation (\ref{vinet:eq4.13}) already provides us with the desired result for $\psi_{0,k+1}$. Using (\ref{vinet:eq2.5a}) in (\ref{vinet:eq4.18a}) with $i=1$, gives similarly:
  \begin{eqnarray} 
  \sqrt{k+1}\,  \psi_{1,k+1} = \ch \rho\, \psi_{0,k} - \sigma \ee^{-i\delta} \psi_{1,k} 
   - \sqrt{2} \ee^{-i\theta} \sh \rho\, \psi_{2,k}.\label{vinet:eq4.19}
  \end{eqnarray}  
Now, when $n=0$, the 5-term recurrence relation   (\ref{vinet:eq2.7}) gives $\psi_{2,k}$ in terms of $\psi_{0,k}$ and $\psi_{1,k}$:
  \begin{eqnarray}   
 \fl -\sqrt{2} \ch \rho \sh \rho \ee^{-i\theta} \psi_{2,k}  = \bigg(k - \frac12 \ch 2\rho + \frac12 - \sigma^2\bigg) \, \psi_{0,k}\nonumber\\
   - \sigma (\ee^{i(\delta-\theta)} \sh \rho - \ee^{-i\delta} \ch \rho)\psi_{1,k}.\label{vinet:eq4.20}
  \end{eqnarray}
Of course, as noted in Section~\ref{vinet:section2}, if $\rho = 0$ we see that $\psi_{1,k}$ is given in terms of $\psi_{0,k}$ and $\psi_{0,k}$ and $\psi_{1,k}$ are thus dependent; similarly, if $\sigma = 0$ $\psi_{2,k}$ is expressed only in terms of $\psi_{0,k}$ and does not depend on $\psi_{1,k}$. Using (\ref{vinet:eq4.20}) to substitute in (\ref{vinet:eq4.19}) for $\psi_{2,k}$ yields after simplification
  \begin{eqnarray} \label{vinet:eq4.21}
  \sqrt{k+1}\,  \psi_{1,k+1} = \frac{1}{\ch \rho}(k+1-\sigma^2) \, \psi_{0,k} - \sigma \ee^{i(\delta - \theta)}  \Th \rho\, \psi_{1,k}.
  \end{eqnarray}
  The same approach, using (\ref{vinet:eq2.5b}) in (\ref{vinet:eq4.18b})
   leads to
    \numparts  \label{vinet:eq4.22}
  \begin{eqnarray} \label{vinet:eq4.22a}
  \sqrt{k}\,  \psi_{0,k-1} = \ch \rho \psi_{1,k} - \sigma \ee^{i\delta} \psi_{0,k}
  \end{eqnarray}   
  for $i=0$, and to
   \begin{eqnarray} \label{vinet:eq4.22b}
  \sqrt{k}\,  \psi_{1,k-1} = - \frac{\ee^{i\theta}}{\sh \rho} (k - \sigma^2) \, \psi_{0,k} - \sigma \ee^{i(\theta - \delta)} \coth \rho\,\psi_{1,k}
    \end{eqnarray}    
    \endnumparts
    for $i=1$ (after using (\ref{vinet:eq4.20})).
    
   At this point, it is useful to return to the formulation  (\ref{vinet:eq2.10}) where the pair $\psi_{0,k}$ and $\psi_{1,k}$ is viewed as the 2-vector $\Psi_{0,k}$. Introducing the $2 \times 2$ $k$-dependent ladder matrices
  \numparts   \begin{eqnarray} \label{vinet:eq23}
    \mathcal{A}_k &=
    \left(\begin{array}{l l}
    -\sigma \ee^{i\delta} &\ch \rho\\
    \frac{\ee^{i\theta}}{\sh \rho} (\sigma^2 - k) &-\sigma \coth \rho\ee^{i(\theta-\delta)}
     \end{array}
  \right)\label{vinet:eq2.23a}\\[2\jot]
    \bar{\mathcal{A}}_k &=
    \left(\begin{array}{l l}
    -\sigma \ee^{-i\delta} &-\ee^{-i\theta}\sh \rho\\
    \frac{1}{\ch \rho} (k+1-\sigma^2) &-\sigma \ee^{i(\delta-\theta)} \Th \rho
     \end{array}
  \right),\label{vinet:eq2.23b}
  \end{eqnarray}
    \endnumparts
        it is checked that the relations (\ref{vinet:eq4.22}) and (\ref{vinet:eq4.13}), (\ref{vinet:eq4.21}) respectively translate into the lowering and raising relations:
          \begin{eqnarray} \label{vinet:eq4.24}
            {\mathcal{A}}_k \Psi_{0,k} = \sqrt{k}\,  \Psi_{0,k-1}, \qquad \bar{\mathcal{A}}_k \Psi_{0,k} = \sqrt{k+1}\,  \Psi_{0,k+1}.
  \end{eqnarray}
  It is also immediate to verify that
  \begin{eqnarray} \label{vinet:eq4.25}
  {\mathcal{A}}_{k+1} \bar{{\mathcal{A}}}_k - \bar{{\mathcal{A}}}_{k-1} {\mathcal{A}}_k = {\mathbbm{1}}_{2\times 2}.
  \end{eqnarray}  
  
    \section{Difference equation} \label{vinet:section5}
  
  We derive in this section the finite-difference equation that the matrix orthogonal polynomials $\mathcal{P}_n(k)$ satisfy. Together with the 3-term recurrence relation 
  (\ref{vinet:eq2.17}), this will express the bispectrality property of these $\mathcal{P}_n(k)$. Recall that (\ref{vinet:eq2.17}) followed from $k \, \psi_{n,k} = \langle k|a^+aS|n\rangle$; expectedly, the difference equation will ``dually'' stem from
  \begin{eqnarray} \label{vinet:eq5.1}  
  n\, \psi_{n,k} = \langle k |S a^+a|n\rangle = \langle k |S a^+aS^{-1}S|n\rangle.
   \end{eqnarray} 
 The transform $Sa^+aS^{-1}$ of $a^+a$ under the automorphism generated by $S$ is evaluated this time using (\ref{vinet:eq4.7}) and its hermitian conjugate. Given that result, following a procedure by now very familiar, we find from (\ref{vinet:eq5.1}):
\begin{eqnarray} \label{vinet:eq5.2}
\fl  n   \, \psi_{n,k} = \lambda (k)   \, \psi_{n,k}  + \sqrt{k}\,  \mu(k)  \,  \psi_{n,k-1} + \sqrt{k+1}\,  \bar{\mu}(k)   \,  \psi_{n,k+1} \nonumber\\
   + \sqrt{k(k-1)}\,  \nu   \, \psi_{n,k-2}  + \sqrt{(k+1)(k+2)}\,  \overline{\nu}\,  \psi_{n,k+2}
  \end{eqnarray}  
  where
 \numparts \label{vinet:eq5.3}
   \begin{eqnarray}
  \lambda(k) &= \bar{\lambda}(k) = \bigg(k + \sigma^2 + \frac12\bigg) \ch 2\rho + \sigma^2 \sh 2\, \rho \cos (2\delta -\theta) - \frac12 \label{vinet:eq5.3a}\\[2\jot]
  \mu &= \sigma (\ch 2\rho \ee^{-i\delta} + \sh 2\, \rho \ee^{i(\delta-\theta)}) \label{vinet:eq5.3b}\\[2\jot]
  \nu &= \frac12 \ee^{-i\theta} \sh 2\, \rho. \label{vinet:eq5.3c}
  \end{eqnarray}
  \endnumparts
  Equation (\ref{vinet:eq5.2}) is easily transcribed in the matrix formalism where
    \begin{eqnarray} 
\Psi_{n,k} =
    \left(\begin{array}{l}
  \psi_{2n,k}\\
 \psi_{2n+1,k}  \end{array}
  \right)= \mathcal{P}_n(k)  \Psi_{0,k}. \label{vinet:eq5.4}
    \end{eqnarray}
One obtains:
  \begin{eqnarray} \label{vinet:eq5.5}
  \fl\lambda(k) \mathcal{P}_n(k)   \Psi_{0,k} 
   + \sqrt{k}\,  \mu \mathcal{P}_n(k-1) \Psi_{0,k-1} + \sqrt{k+1}\,  \bar{\mu} \mathcal{P}_n(k+1) \Psi_{0,k+1} \nonumber\\
  + \sqrt{k(k-1)}\,  \nu \mathcal{P}_n(k-2) \Psi_{0,k-2} \nonumber\\
   + \sqrt{(k+1)(k+2)}\,  \overline{\nu} \mathcal{P}_n(k+2) \Psi_{0,k+2} = \gamma (n) \mathcal{P}_n(k) \Psi_{0,k}
  \end{eqnarray}
  where
    \begin{eqnarray} \label{vinet:eq5.6}
   \gamma(n) =
    \left(\begin{array}{lc c}
  2n &0\\
 0 &2n+1  \end{array}
  \right).
  \end{eqnarray}  
 Here one uses the lowering and raising matrices ${\mathcal{A}}_k$ and $\bar{\mathcal{A}}_k$ of the last section and the ladder relations (\ref{vinet:eq4.24}), to write (\ref{vinet:eq5.5}) in a form where all the (matrix) terms in the equation act on the 2-vector $\Psi_{0,k}$ which can then be factored out since it has no annihilator when $\rho \ne 0 \ne \sigma$ as already discussed. This then gives the following equation for the matrix orthogonal polynomials $\mathcal{P}_n(k)$:
  \begin{eqnarray} \label{vinet:eq5.7}
\fl  \lambda(k) \mathcal{P}_n(k)  + \mu \mathcal{P}_n(k-1) {\mathcal{A}}_k + \bar{\mu} \mathcal{P}_n (k+1) \bar{{\mathcal{A}}}_k\nonumber\\
  + \nu \mathcal{P}_n(k-2) {\mathcal{A}}_{k-1} {\mathcal{A}}_k + \bar{\nu} \mathcal{P}_n(k+2)\bar{{\mathcal{A}}}_{k+1} \bar{{\mathcal{A}}}_k = \gamma (n) \mathcal{P}_n (k).
  \end{eqnarray} 
 Introduce the shift operators $T_{\pm}$ acting on the left as follows on functions of $f$ of the discrete variable $k$:
  \begin{eqnarray} \label{vinet:eq5.8}
f(k) T_{\pm} =  f(k \pm 1).
   \end{eqnarray}  
  Let now $\overset{\leftarrow}{\Delta}_k$ be the right-handed matrix-difference operator defined by
  \begin{eqnarray} \label{vinet:eq5.9}  
 \overset{\leftarrow}{\Delta}_k = \nu T_-^2 {\mathcal{A}}_{k-1}{\mathcal{A}}_k + \bar{\nu} T_+^2 \bar{{\mathcal{A}}}_{k+1} \bar{{\mathcal{A}}}_k + \mu T_- {\mathcal{A}}_k + \bar{\mu} T_+ \bar{{\mathcal{A}}}_k + \lambda (k),
  \end{eqnarray}  
  we see that the matrix orthogonal polynomials $\mathcal{P}_n(k)$ satisfy the difference equation:
    \begin{eqnarray} \label{vinet:eq5.10-1}  
  \mathcal{P}_n(k)  \overset{\leftarrow}{\Delta}_k = \gamma(n)   \mathcal{P}_n(k).
  \end{eqnarray}
    It is straightforward to check that the operator $\overset{\leftarrow}{\Delta}_k$ is hermitian $\overset{\leftarrow}{\Delta}{}_k^+ = \overset{\leftarrow}{\Delta}_k$. Recall that the inner product on the space of $2\times 2$ matrix functions of $k$ is defined by
    \begin{eqnarray} \label{vinet:eq5.11-1}    
\big( \mathcal{P}(k), \mathcal{Q}(k)\big)= \sum_{k=0}^{\infty} \mathcal{P}(k) W(k) \mathcal{Q}^+(k).
  \end{eqnarray}
  The hermitian conjugate of an operator $X$ on that space is hence defined by $(\mathcal{P}, \mathcal{Q}X) = (\mathcal{P}X^+,\mathcal{Q})$. Since $\lambda(k)$ is real, to prove that $\overset{\leftarrow}{\Delta}{}_k^+ = \overset{\leftarrow}{\Delta}_k$, it suffices to show that the two pairs of operators $T_- {\mathcal{A}}_k$ and $T_+ \bar{{\mathcal{A}}}_k$ as well as $T_-^2 {\mathcal{A}}_{k-1}{\mathcal{A}}_k$ and $T_+^2 \bar{{\mathcal{A}}}_{k+1} \bar{{\mathcal{A}}}_k$ are both hermitian conjugate pairs. Let us look at the first. We have
    \begin{eqnarray}
(\mathcal{P}(k), \mathcal{Q}(k)T_- {\mathcal{A}}_k) 
&= \sum_{k=0}^{\infty}   \mathcal{P}(k) \mathcal{W}(k)(\mathcal{Q}(k-1){\mathcal{A}}_k)^+\nonumber\\
&= \sum_{k=0}^{\infty}\mathcal{P}(k)\Psi_{0,k}\Psi^+_{0,k} {\mathcal{A}}_k^+ \mathcal{Q}(k-1)^+\nonumber\\
  &= \sum_{k=0}^{\infty}\mathcal{P}(k)\sqrt{k}\,  \Psi_{0,k}\Psi^+_{0,k-1} \mathcal{Q}(k-1)^+ \nonumber\\
  &= \sum_{k=0}^{\infty}\mathcal{P}(k+1) \sqrt{k+1}\,  \Psi_{0,k+1}\Psi^+_{0,k}  \mathcal{Q}(k)^+\nonumber\\
  &= \sum_{k=0}^{\infty}\mathcal{P}(k) T_+ \bar{{\mathcal{A}}}_k \Psi_{0,k}\Psi_{0,k}^{+} \mathcal{Q}(k)^+\nonumber\\
  &= (\mathcal{P}(x) T_+ \bar{{\mathcal{A}}}_k, \mathcal{Q}(k))  \label{vinet:eq5.10}      
  \end{eqnarray}
    where we used (\ref{vinet:eq3.5}), (\ref{vinet:eq4.24}) and (\ref{vinet:eq5.8}), which confirms that
     \begin{eqnarray} \label{vinet:eq5.11}  
    (T_- {\mathcal{A}}_k)^+ = T_+ \bar{{\mathcal{A}}}_k.
  \end{eqnarray}    
One demonstrates similarly that    
    \begin{eqnarray} \label{vinet:eq5.12}  
    (T_-^2 {\mathcal{A}}_{k-1}{\mathcal{A}}_k)^+ = T_+^2 \bar{{\mathcal{A}}}_{k+1} \bar{{\mathcal{A}}}_k
  \end{eqnarray}        
    to complete the proof that $\overset{\leftarrow}{\Delta}{}_k^+ = \overset{\leftarrow}{\Delta}_k$.
    
        \section{Ladder operators and Rodrigues' formula} \label{vinet:section6}
    
    Right-handed lowering and raising operators $\overset{\leftarrow}{L}_k$ and  $\overset{\leftarrow}{R}_k$ can further be obtained for the matrix polynomials $\mathcal{P}_n(k)$. They are found to act as follows:
     \numparts  \label{vinet:eq6.1}
  \begin{eqnarray}  
  \mathcal{P}_n(k) \overset{\leftarrow}{L}_k &=  \Theta_n \mathcal{P}_{n-1}(k),\label{vinet:eq6.1a}\\
    \mathcal{P}_n(k) \overset{\leftarrow}{R}_k &=  \Theta_{n+1} \mathcal{P}_{n+1}(k)\label{vinet:eq6.1b}
    \end{eqnarray}    
    \endnumparts
    where
  \begin{eqnarray} \label{vinet:eq6.2}
  \Theta_n =
\left(\begin{array}{c c} 
\sqrt{2n(2n-1)} & 0\\[2\jot]
0 & \sqrt{2n(2n+1)}  
  \end{array}
  \right).    
  \end{eqnarray}    
In order to determine the expressions of  $\overset{\leftarrow}{L}_k$ and  $\overset{\leftarrow}{R}_k$, we respectively start from the identities
     \numparts  \label{vinet:eq6.3}
  \begin{eqnarray}  
  \sqrt{n(n-1)}\,  \langle k |S|n-2\rangle = \langle k |Sa^2|n\rangle = \langle k |Sa^2 S^{-1}S|n\rangle \label{vinet:eq6.3a}
  \end{eqnarray}
  and
   \begin{eqnarray}  
   \sqrt{(n+1)(n+2)} \,  \langle k |S|n + 2\rangle = \langle k |S(a^+)^2 |n\rangle = \langle k |S(a^+)^2 S^{-1} S |n. \rangle\label{vinet:eq6.3b}
  \end{eqnarray}
 \endnumparts

We here limit ourselves to a brief summary of the derivation since the steps are analogous to those followed to obtain the difference equation. We appeal again to (\ref{vinet:eq4.7}) and its hermitian conjugate the compute $Sa^2S^{-1}$ and $S(a^+)^2S^{-1}$. The results are then used in (\ref{vinet:eq6.3}) to obtain the corollary relations between $\psi_{n\pm 2,k}$ and $\psi_{n,k}$, $\psi_{n,k\pm1}$ and $\psi_{n,k\pm 2}$. One then translates these relations in terms of the 2-vectors $\Psi_{n,k}$. As in Section~\ref{vinet:section5}, one finally employs the ladder matrices ${\mathcal{A}}_k$ and $\bar{\mathcal{A}}_k$ to get identities for the matrix polynomials $\mathcal{P}_n(k)$ by eliminating $\Psi_{0,k}$. This brings one to find that :
 \begin{eqnarray}    
\fl    \overset{\leftarrow}{R}_k
     = \frac12 (\ch 2\rho + 1) T_-^2  {{\mathcal{A}}}_{k-1}{\mathcal{A}} _k      + \frac12 \ee^{2i\theta} (\ch 2\rho -1) T_+^2 {\bar{{\mathcal{A}}}}_{k+1} {\bar{{\mathcal{A}}}}_k\nonumber\\
     + \sigma [\ee^{i\delta} (\ch 2\rho + 1) + \ee^{-i(\delta-\theta)} \sh 2\, \rho\,] T_-{{\mathcal{A}}}_k\nonumber\\
     + \sigma [ \ee^{i(\delta+\theta)}\sh 2\, \rho + \ee^{-i(\delta -2\theta)}(\ch 2\rho - 1)] T_+ {\bar{{\mathcal{A}}}}_k\nonumber\\
     + \bigg[ \ee^{i\theta} \sh 2\, \rho \bigg(k + \sigma^2 + \frac12\bigg) + \frac{\sigma^2}{2} \ch 2\rho\, (\ee^{2i\delta} + \ee^{-2i(\delta-\theta)})\nonumber\\
    \qquad + \frac{\sigma^2}{2} (\ee^{2i\delta} - \ee^{-2i(\delta-\theta)})\bigg] \label{vinet:eq6.4}
\end{eqnarray}
and
  \begin{eqnarray} \label{vinet:eq6.5}
  \overset{\leftarrow}{L}_k = (\overset{\leftarrow}{R}_k)^+
  \end{eqnarray}
where one should remember (\ref{vinet:eq5.11}) and (\ref{vinet:eq5.12}).

Note now from formula (\ref{vinet:eq6.1b}), that the repeated action of $\overset{\leftarrow}{R}_k$ on the constant matrix ${\mathbbm{1}}_{2\times 2}$ gives
  \begin{eqnarray} \label{vinet:eq6.6}
{\mathbbm{1}} \cdot  (\overset{\leftarrow}{R}_k)^n = \left( \prod_{i=1}^n \Theta_i\right) \mathcal{P}_n(k).
 \end{eqnarray}
 This is the Rodrigues' formula for the matrix polynomials $ \mathcal{P}_n(k)$.

\section{Decomposition in terms of Charlier and Meixner polynomials}\label{vinet:section7}

In this section we shall exploit the product structure of $S$ to show that classical scalar orthogonal polynomials in a discrete variables are building blocks of the matrix orthogonal polynomials $\mathcal{P}_n(k)$. The Charlier and Meixner polynomials will hence be seen to enter the picture.

Define the matrix elements of the two factors in $S$ (\ref{vinet:eq1.16}):
  \begin{eqnarray}
  \chi_{m,k}&= \langle k | \ee^{(va - \bar{v} a^+)} |m\rangle  \label{vinet:eq7.1}\\
    \varphi_{n,m}&= \langle m | \ee^{[wa^2-\overline{w}(a^+)^2]/2} |n\rangle. \label{vinet:eq7.2}
  \end{eqnarray}
  In terms of these we have
   \begin{eqnarray} 
   \, \psi_{n,k} &= \langle k |S|n\rangle = \langle k |\ee^{(va-\bar{v}a^+)}\ee^{[wa^2 - \overline{w}(a^+)^2]/2} |n \rangle\nonumber \\
   &=\sum_{m=0}^{\infty} \langle k | \ee^{(va - \bar{v} a^+)} |m\rangle \langle m | \ee^{ [wa^2 -\bar w(a^+)^2]/2}|n\rangle \nonumber\\
   &= \sum_{m=0}^{\infty} \chi_{m,k} \varphi_{n,m}. \label{vinet:eq7.3}
     \end{eqnarray}
     
     We shall now calculate the matrix elements $\chi_{m,k}$ and $\varphi_{n,m}$. Note that this will provide by the same token, the matrix elements of $S$ when $\rho = 0$ or $\sigma = 0$. Indeed it is obvious from (\ref{vinet:eq7.3}) that
 \numparts \label{vinet:eq7.4}
  \begin{eqnarray} \label{vinet:eq7.4a}
     \, \psi_{n,k} = \chi_{n,k} \quad { \rm when} \quad w = 0, \qquad (\rho = 0)
  \end{eqnarray}
  and that
    \begin{eqnarray}  
     \, \psi_{n,k} = \varphi_{n,k} \quad { \rm when} \quad v = 0, \qquad (\sigma = 0).\label{vinet:eq7.4b}
  \end{eqnarray}
  \endnumparts
  These will again be determined from their recurrence relation.
  
  \subsection{The matrix element $\chi_{n,k}$: Charlier polynomials}\label{vinet:subsection7.1}
  
Using   (\ref{vinet:eq7.4a}), the recurrence relation for $\chi_{n,k}$ is obtained from  (\ref{vinet:eq2.7}) by letting $\rho = 0$. One observes that it then only has three terms and that it reads :
   \begin{eqnarray} \label{vinet:eq7.5}
k \chi_{n,k} = (n + \sigma^2) \chi_{n,k}  - \sqrt{n} \,  \sigma \ee^{-i\delta} \chi_{n-1,k} - \sqrt{n+1}\, \sigma \ee^{-i\delta} \chi_{n+1,k}.
   \end{eqnarray} 
   An examination of (\ref{vinet:eq7.5}) shows naturally that its solutions are of the form  $\chi_{n,k} = P_n(x) \chi_{0,k}$ where $P_n(k)$ is a polynomial of degree $n$ in the variable $k$. If we let
   \begin{eqnarray} \label{vinet:eq7.6}   
P_n(k)= \frac{1}{\sqrt{n!}} (-\sigma \ee^{-i\delta})^{-n}  \widehat{P}_n(k),   
     \end{eqnarray} 
     we see that the normalized polynomials $\widehat{P}_n(k)$ obey:
    \begin{eqnarray} \label{vinet:eq7.7}  
    \widehat{P}_{n+1}(k) = (k-n-\sigma^2) \widehat{P}_n(k) - n \sigma^2 \widehat{P}_{n-1}(k).
    \end{eqnarray}
    Comparing with  (\ref{vinet:eqA.9}), we find that  $ \widehat{P}_n(k)$ are normalized Charlier polynomials with $a = \sigma^2$. Reverting to standard Charlier polynomials with the help of (\ref{vinet:eqA.10}), we thus have
       \begin{eqnarray} \label{vinet:eq7.8}   
\widehat{P}_n(k) = (-\sigma^2)^n C_n(k;\sigma^2),
     \end{eqnarray} 
     and
       \begin{eqnarray} \label{vinet:eq7.9}      
 \chi_{n,k} = \frac{1}{\sqrt{n!}} (\sigma \ee^{i\delta})^n C_n(k;\sigma^2)  \chi_{0,k}.     
      \end{eqnarray}     
     
 There remains to evaluate $\chi_{0,k} =  \langle k | \ee^{va  - \bar{v}a^+}|0\rangle$ which is done directly with the help of the coherent states properties. From (\ref{vinet:eq4.1}), one has (similarly to (\ref{vinet:eq4.2})):
  \begin{eqnarray} \label{vinet:eq7.10}      
  \ee^{(va-\bar{v}a^+)} = \ee^{ v\bar{v}/2} \ee^{va} \cdot \ee^{-\bar{v}a^+}.
       \end{eqnarray}  
 Hence,      
  \begin{eqnarray} \label{vinet:eq7.11}      
  \ee^{va-\bar{v}a^+}|0\rangle  = \ee^{v\bar{v}/2} \ee^{va}|-\bar{v}\rangle =    \ee^{-v \bar{v} /2}|-\bar{v} \rangle .
       \end{eqnarray}         
 using successively (\ref{vinet:eq1.11a}) and (\ref{vinet:eq1.11b}). One then obtains
   \begin{eqnarray} \label{vinet:eq7.12}        
 \chi_{0,k}
 &= \langle k | \ee^{va - \bar{v} a^+} |0\rangle = \ee^{-v\bar{v}/2} \langle k |-\bar{v}\rangle \nonumber\\[2\jot]
 &= \frac{(-\bar{v})^k}{\sqrt{k!}} \ee^{-v\bar{v}/2} = \frac{\ee^{-\sigma^2/2}}{\sqrt{k!}} (-\sigma \ee^{-i\delta})^k      
        \end{eqnarray}            
with the help of (\ref{vinet:eq1.11c}) and $v = \sigma \ee^{i\delta}$. Collecting gives the following formula for the matrix elements $\chi_{n,k}$:
\begin{eqnarray} \label{vinet:eq7.13}          
        \chi_{n,k} = \frac{(-1)^k}{\sqrt{n!k!}} \sigma^{n+k} \ee^{i\delta(n-k)} \ee^{-\sigma^2/2} C_n(k;\sigma^2)
        \end{eqnarray}       
        in terms of Charlier polynomials.       

\subsection{The matrix elements $\varphi_{n,k}$: Meixner polynomials} \label{vinet:subsection7.2}

As per  (\ref{vinet:eq7.4b}), the recurrence relation for $\varphi_{n,k}$ is the special case of (\ref{vinet:eq2.7}) when $\sigma = 0$, that is:
\begin{eqnarray}
\fl  k \varphi_{n,k} 
 = \bigg[ \bigg(n+ \frac12\bigg) \ch 2\rho - \frac12 \bigg] \varphi_{n,k}  - \frac12 \sqrt{n(n-1)} \sh  2\rho \ee^{i\theta} \varphi_{n-2,k}\nonumber\\[2\jot]
 -\frac12 \sqrt{(n+1)(n+2)} \,  \sh 2\, \rho \ee^{-i\theta} \varphi_{n+2,k}.
\label{vinet:eq7.14}   
         \end{eqnarray}    
It has 3-terms with increments of 2. The elements with even and odd $n$ are decoupled and the solutions are clearly of the form\footnote{In this subsection, as in the preceding ones, the use of $P_n$ (and $Q_n$) to denote polynomials is generic. The identifications do not carry from one section to another---we trust there should be no confusion.}
 \numparts
 \begin{eqnarray} \label{vinet:eq7.15}
\varphi_{2n,k} &= P_n(k) \varphi_{0,k}, \label{vinet:eq7.15a}\\
\varphi_{2n+1,k} &= Q_n(k) \varphi_{1,k}. \label{vinet:eq7.15b}
        \end{eqnarray}   
\endnumparts
Substitution in  (\ref{vinet:eq7.14}) gives for $P_n$ and $Q_n$ the following recursions:
 \numparts
  \begin{eqnarray}      \label{vinet:eq7.16}   
\fl  kP_n(k)
  = \bigg[\bigg(2n +\frac12\bigg) \ch 2\rho - \frac12\bigg] P_n(k)
   -\frac12 \sqrt{2n(2n-1)} \sh 2\, \rho \ee^{i\theta} P_{n-1}(k)\nonumber\\
   -\frac12 \sqrt{(2n+1)(2n+2)} \sh 2\, \rho \ee^{-i\theta} P_{n+1}(k)   \label{vinet:eq7.16a}   \\ \nonumber\\
  kQ_n(k)
  = \bigg[\bigg(2n +\frac32\bigg) \ch 2\rho - \frac12\bigg] Q_n(k)\nonumber\\
   -\frac12 \sqrt{2n(2n+1)} \sh 2\, \rho \ee^{i\theta}Q_{n-1}(k)\nonumber\\
   -\frac12 \sqrt{(2n+2)(2n+3)} \sh 2\, \rho \ee^{-i\theta} Q_{n+1}(k).   \label{vinet:eq7.16b}       
          \end{eqnarray}  
          \endnumparts  
 Let us first focus on $P_n(k)$. We can normalize  (\ref{vinet:eq7.16a}) by setting
  \begin{eqnarray}      \label{vinet:eq7.17}   
  P_n(k) = \frac{1}{\sqrt{2n!}} \bigg( -\frac14 \ee^{-i\theta} \sh 2\, \rho\bigg)^{-n} \widehat{P}_n(k),
          \end{eqnarray}  
to find that (\ref{vinet:eq7.16a}) reduces to
\begin{eqnarray}      \label{vinet:eq7.18}             
\fl  \widehat{P}_{n+1}(k)
   =  \bigg[ \frac{k}{2} - \bigg( n + \frac14\bigg) \ch 2\rho + \frac14\bigg] \widehat{P}_n(k) \nonumber\\
   -\frac14\bigg[n\bigg(n-\frac12\bigg)\bigg] \sh^2 2\rho\, \widehat{P}_{n-1}(k).
 \end{eqnarray}
 With $c$, a parameter defined by
  \begin{eqnarray} \label{vinet:eq7.19}     
 c = \Th^2 \rho \qquad 0 \le c < 1
  \end{eqnarray}
  so that
  \begin{eqnarray}      \label{vinet:eq7.20}     
 \ch 2\rho = \frac{1+c}{1-c}; \qquad \sh^2 2\rho = \frac{4c}{(1-c)^2},
  \end{eqnarray}  
eq.~(\ref{vinet:eq7.18}) can be rewritten as follows:
  \begin{eqnarray}      \label{vinet:eq7.21}     
\fl  \widehat{P}_{n+1}(k) =\left[\frac{k}2 - \left( \frac{n+c(n+1/2)}{1-c}\right)\right] \widehat{P}_n(k) \nonumber\\[2\jot]
 - n(n-1/2) \frac{c}{(1-c)^2} \widehat{P}_{n-1}(x).
  \end{eqnarray}
Comparing with (\ref{vinet:eqA.15}), we conclude that $\widehat{P}_n(k)$ coincide with the normalized Meixner polynomials with $x = k/2$, $\beta = 1/2$ and $c$ as is. Using (\ref{vinet:eqA.16}), we thus have in terms of the standard Meixner polynomials:
  \begin{eqnarray}      \label{vinet:eq7.22}    
  \widehat{P}_n(k) = \bigg(\frac12\bigg)_n \bigg( \frac{c}{c-1}\bigg)^n M_n \bigg(\frac{k}2; \frac12, c\bigg).
  \end{eqnarray}          
  Recalling that $\big(\frac12\big)_n = \frac12 \cdot \frac32 \ldots \big(\frac12 + n -1\big)$, we verify that
    \begin{eqnarray}      \label{vinet:eq7.23}   
 \bigg( \frac12\bigg)_n  = \frac{1}{2^{2n}} \frac{(2n)!}{n!}.
  \end{eqnarray}
Putting (\ref{vinet:eq7.17}), (\ref{vinet:eq7.22}), and (\ref{vinet:eq7.23}) together, we get after some simplification (with the help of (\ref{vinet:eq7.19}) and (\ref{vinet:eq7.20})), that
    \begin{eqnarray}      \label{vinet:eq7.24}   
P_n(k) = \frac{1}{2^n} \frac{\sqrt{(2n)!}}{n!} \ee^{in\theta} \Th^n\rho\, M_n\bigg(\frac{k}2; \frac12, \Th^2\rho\bigg).
 \end{eqnarray}  
The polynomials of  (\ref{vinet:eq7.15b}) and (\ref{vinet:eq7.16b}) are obtained similarly. Their recurrence relation is normalized by taking
  \begin{eqnarray} \label{vinet:eq7.25}
Q_n(k) = \bigg( -\frac14 \ee^{-i\theta} \sh 2\, \rho \bigg)^{-n} \frac{1}{\sqrt{(2n+1)!}} \widehat{Q}_n(k), 
 \end{eqnarray}  
it then becomes
   \begin{eqnarray} \label{vinet:eq7.26}
\fl  \widehat{Q}_{n+1}(k) =  \bigg[ \frac{k}{2} - \bigg(n + \frac{3}{4}\bigg) \ch 2\rho + \frac14\bigg] \widehat{Q}_n(k)\nonumber \\[2\jot]
 -n \bigg(n +  \frac12\bigg) \frac{\sh^2\rho}{4} \widehat{Q}_{n-1}(k).
 \end{eqnarray}  
Using the same parameter $c$ as in (\ref{vinet:eq7.19}) and (\ref{vinet:eq7.20}), after simple algebra, one casts (\ref{vinet:eq7.26}) in the form
\begin{eqnarray} \label{vinet:eq7.27}
\fl  \widehat{Q}_{n+1}(k)
 =  \bigg( \frac{k-1}{2} - \frac{[n+(n+3/2)c]}{1-c}\bigg) \widehat{Q}_n(k)\nonumber\\[2\jot]
 -n \bigg( n + \frac12\bigg) \frac{c}{(1-c)^2} \widehat{Q}_{n-1} (k).
   \end{eqnarray}        
 Comparing again with (\ref{vinet:eqA.15}), we realize that in this case the $\widehat{Q}_n(k)$ are normalized Meixner polynomials with $x = (k-1)/{2}$, $\beta = 3/2$ and $c$ unchanged. Passing to the standard Meixner polynomials with the help of (\ref{vinet:eqA.16}), we therefore have
  \begin{eqnarray} \label{vinet:eq7.28}  
 \widehat{Q}_n(k) = \bigg(\frac32\bigg)_n \bigg( \frac{c}{c-1} \bigg)^n M_n \bigg( \frac{k-1}{2}; \frac32, c \bigg)
   \end{eqnarray}      
 where
   \begin{eqnarray} \label{vinet:eq7.29}
\bigg( \frac32\bigg)_n = \frac32 \cdot \frac52 \ldots \bigg(\frac32 + n -1 \bigg) = \frac{1}{2^{2n}} \frac{(2n+1)!}{n!}.
   \end{eqnarray}     
Bringing the pieces together ((\ref{vinet:eq7.25}), (\ref{vinet:eq7.28}), (\ref{vinet:eq7.29})) with $c = \Th^2\rho$  still, gives
  \begin{eqnarray} \label{vinet:eq7.30}
Q_n(k) = \frac{1}{2^n} \frac{\sqrt{(2n+1)!}}{n!} \ee^{in\theta} \Th^n\rho\, M_n \bigg( \frac{k-1}{2}; \frac32 , \Th^2\rho\bigg).
   \end{eqnarray}               
   Keeping  (\ref{vinet:eq7.15a}, \ref{vinet:eq7.15b}) in mind, there now only remains to compute $\varphi_{0,k}$ and $\varphi_{1,k}$:
     \begin{eqnarray} \label{vinet:eq7.31}
   \varphi_{i,k} = \langle k | \ee^{[wa^2 - \overline{w}(a^+)^2]/2}|i\rangle, \qquad i = 0,1,
   \end{eqnarray}  
 to complete the determination of the matrix element $\varphi_{n,k}$. Let us point out that $\varphi_{0,k}$ is the overlap of Perelomov $\mathrm{SU}(1,1)$ coherent state  \cite{vinet:ref7} with $k$. We need to look at the two cases $i=0,1$ separately. Using the disentangling theorem (4.4) :
      \begin{eqnarray} \label{vinet:eq7.32}
 \ee^{[wa^2 - \overline{w}(a^+)^2]/2}|0\rangle &= (\ch \rho)^{-1/2} \exp\bigg[-\frac12 \ee^{-i\theta}  \Th \rho\, (a^+)^2\bigg] | 0\rangle\nonumber\\[2\jot]
 &= (\ch \rho)^{-1/2} \sum_{n=0}^{\infty} \frac{\sqrt{(2n)!}}{n!} \bigg( - \frac12 \ee^{-i\theta}  \Th \rho\bigg)^n |2n \rangle.
 \end{eqnarray}  
   Consequently,
       \begin{eqnarray} \label{vinet:eq7.33}  
   \varphi_{0,2k} &= (\ch \rho)^{-1/2} \frac{\sqrt{(2k)!}}{k!} \bigg( -\frac12 \ee^{-i\theta} \Th \rho\bigg)^k\nonumber\\
   \varphi_{0,2k+1} &= 0.
   \end{eqnarray}   
Similarly for $i=1$,
\begin{eqnarray} \label{vinet:eq }
&\ee^{[wa^2 - \overline{w}(a^+)^2]/2}|1\rangle 
 = (\ch \rho)^{-3/2}  \exp\bigg[ -\frac12 \ee^{-i\theta}  \Th \rho\, (a^+)^2\bigg] | 1\rangle\nonumber\\[2\jot]
&=(\ch \rho)^{-3/2} \sum_{n=0}^{\infty} \frac1{n!} \bigg( -\frac12 \ee^{-i\theta} \Th \rho\bigg)^n (a^+)^{2n} \cdot a^+ |0\rangle\nonumber\\[2\jot]
&= (\ch \rho)^{-3/2} \sum_{n=0}^{\infty}   \frac{\sqrt{(2n+1)!}}{n!} \bigg( - \frac12 \ee^{-i\theta}  \Th \rho\bigg)^n |2n+1\rangle \label{vinet:eq7.34}
   \end{eqnarray}  
from where follows that
\begin{eqnarray} \label{vinet:eq7.35}
\varphi_{1,2k} &=0,\nonumber\\
\varphi_{1,2k+1} &= (\ch \rho)^{-3/2} \frac{\sqrt{(2k+1)!}}{k!} \bigg( - \frac12 \ee^{-i\theta}  \Th \rho \bigg)^k.
   \end{eqnarray}     
   In summary, given (\ref{vinet:eq7.15a}, \ref{vinet:eq7.15b}) and in view of (\ref{vinet:eq7.33}) and (\ref{vinet:eq7.35}), the only non-zero $\varphi_{n,k}$ are the even-even and odd-odd elements. Their fomulas are found to be
 \numparts
\begin{eqnarray} \label{vinet:eq7.36}   
   \varphi_{2n,2k} 
   &= \frac{(-1)^k}{2^{k+n}} \frac{\sqrt{(2k)!(2n)!}}{k!n!} \ee^{i(n-k)\theta}\frac{( \Th \rho)^{k+n}}{(\ch \rho)^{1/2}} 
    \cdot M_n\bigg(k; \frac12, \Th^2\rho\bigg)\label{vinet:eq7.36a}   
 \\[2\jot]
   \varphi_{2n+1, 2k+1} 
   &= \frac{(-1)^k}{2^{n+k}} \frac{\sqrt{(2k+1)!(2n+1)!}}{k!n!} \ee^{i(n-k)\theta}\frac{( \Th \rho)^{k+n}}{(\ch \rho)^{3/2}} \nonumber\\
    &\qquad \cdot M_n\bigg(k; \frac32, \Th^2\rho\bigg)\label{vinet:eq7.36b}   
 \end{eqnarray}  
\endnumparts
   when combining  (\ref{vinet:eq7.24}) with (\ref{vinet:eq7.33}) and with  (\ref{vinet:eq7.35}).
   
   \subsection{The convolution for the general matrix elements}\label{vinet:subsection7.4}
   
   The decomposition for the general matrix elements is obtained by bringing together the results of the preceding subsections. From  (\ref{vinet:eq7.3}),  (\ref{vinet:eq7.33}) and  (\ref{vinet:eq7.35}) we have that
 \begin{eqnarray} \label{vinet:eq7.37}   
 \Psi_{n,k} = 
   \left(\begin{array}{c}
  \psi_{2n,k} \\
  \psi_{2n+1,k}
  \end{array}
  \right) = \sum_m
    \left(\begin{array}{c}
  \chi_{2m,k}\, \varphi_{2n,2m} \\
  \chi_{2m+1,k}\, \varphi_{2n+1,2m+1}
  \end{array}
  \right).
  \end{eqnarray}
  Combining   (\ref{vinet:eq7.13}) and  (\ref{vinet:eq7.36}), we find after simple algebraic simplifications
  \begin{eqnarray} \label{vinet:eq7.38}   
\fl \Psi_{n,k} = \mathcal{P}_n(k) \Psi_{0,k}
 = \frac{(-1)^k}{\sqrt{k!}} \frac{\sigma^k}{2^nn!} \ee^{-\sigma^2/2} \ee^{i(n\theta - k \delta)}  \Th^n \rho\nonumber \\[2\jot]
  \cdot \sum_{m=0}^{\infty} \frac{(-\sigma^2 \ee^{i(2\delta-\theta)} \Th \rho)^m}{2^mm!}\nonumber\\[2\jot]
   \left(\begin{array}{l}
\frac{\sqrt{(2n)!}}{(\ch \rho)^{1/2}} C_{2m}(k; \sigma^2) M_n\big( m; \frac12; \Th^2\rho\big)\\[3\jot]
\frac{\sqrt{(2n+1)!}}{(\ch \rho)^{3/2}} \sigma \ee^{i\delta} C_{2m+1} (m,\sigma^2) M_n\big(m;\frac32;\Th^2\rho\big)
     \end{array}
  \right).
   \end{eqnarray}  
   
   \subsection{$\Psi_{0,k}$ revisited} \label{vinet:subsection7.5}
   
   It is instructive to examine how the expressions for $\psi_{0,k}$ and $\psi_{1,k}$ that we obtained in Section~\ref{vinet:section4} through recurrence relations can be recovered from the expansion (\ref{vinet:eq7.38}) in terms of Charlier and Meixner polynomials. As we shall see, it requires the use of formulas obtained by Gessel~\cite{vinet:ref14}, that are not widely known, for sums of Charlier polynomials. 
   
   For notational ease, let
     \begin{eqnarray} \label{vinet:eq7.39}
     t = \ee^{i(2\delta - \theta)} \Th \rho.
  \end{eqnarray}
   Since $M_0(x;\beta, c) = 1$, from (\ref{vinet:eq7.38}) we have
  \begin{eqnarray} 
\fl   \Psi_{0,k} = 
  \left(\begin{array}{l}
  \psi_{0,k}\\
     \psi_{1,k}
  \end{array}
  \right) =
  \frac{1}{\sqrt{k!}} (-\sigma \ee^{-i\delta})^k \ee^{-\sigma^2/2} \nonumber\\
  \cdot \sum_{m=0}^{\infty} \frac{1}{m!} \left(\frac{-\sigma^2t}{2}\right)^m
    \left(\begin{array}{l}
\frac{1}{(\ch \rho)}^{1/2} C_{2m} (k; \sigma^2)\\[2\jot]
     \frac{\sigma \ee^{i\delta}}{(\ch \rho)^{3/2}} C_{2m+1} (k; \sigma^2)\label{vinet:eq7.40}
      \end{array}
  \right).
   \end{eqnarray}   
   
The well-known generating function (\ref{vinet:eqA.10}) gives $\sum_{m=0}^{\infty} C_m(x;a) t^m/m!$. Here however, the degree of the polynomials in the sums are $2m$ or $2m+1$. Fortunately, the needed results have been worked out and are recorded in the Appendix. We shall have to treat the two components of (\ref{vinet:eq7.40}) separately. Consider first the top one. Using (\ref{vinet:eqA.12}), we find
  \begin{eqnarray} \label{vinet:eq7.41}
 \fl \psi_{0,k} =
   \frac{1}{\sqrt{k}\, } (-\sigma \ee^{-i\delta})^k \ee^{-\sigma^{2/2}} \cdot \frac{1}{(\ch \rho)^{1/2}}\nonumber\\[2\jot]
   \cdot \ee^{-\sigma^2t/2} \sum_{\ell=0}^{[k/2]} \frac{(-k)_{2\ell}}{(1+t)^{2\ell - k}} \frac{(-1)^{\ell}}{\ell !}\left( \frac{t}{2\sigma^2}\right)^{\ell},
   \end{eqnarray}   
   where $[k/2]$ denotes the integer part of $k/2$.
   
   Noting that
   \begin{eqnarray} \label{vinet:eq7.42}
   (-k)_{\ell} = (-k)(-k+1) \ldots (-k + 2\ell -1) = (-1)^{2\ell} \frac{k!}{(k-2\ell)!}
    \end{eqnarray}   
and writing
  \begin{eqnarray} \label{vinet:eq7.43}
\left(\frac{t}{2\sigma^2}\right)^{\ell} = \left( \frac{\sqrt{2}\sigma}{t^{1/2}}\right)^{k-2\ell} \cdot \frac{t^{k/2}}{2^{k/2}\sigma^k}  
   \end{eqnarray}
   we see that (\ref{vinet:eq7.41}) can be cast in the form
    \begin{eqnarray} \label{vinet:eq7.44}
 \fl  \psi_{0,k} = 
    \frac{1}{\sqrt{k!}} \frac{\ee^{-ik\delta}}{(\ch \rho)^{1/2}} \ee^{-\sigma^2(1+t)/2} \left( \frac{t}{2} \right)^{k/2}\nonumber \\[2\jot]
    \cdot \left[ k! \sum_{\ell=0}^{[k/2]} \frac{(-1)^{\ell}}{\ell ! (k-2\ell)!} \left( -\frac{2\sigma}{\sqrt{2}} (t^{1/2} + t^{-1/2}) \right)^{k-2\ell} \right].
    \end{eqnarray}  
   From the  explicit expression (\ref{vinet:eqA.3}) for the Hermite polynomials, we observe that the expression in square brackets in (\ref{vinet:eq7.44}) is actually equal to $H_k(s)$ where
    \begin{eqnarray} \label{vinet:eq7.45}     
s = -\frac{\sigma}{\sqrt{2}} (t^{1/2} + t^{-1/2})
   \end{eqnarray}  
is exactly the variable defined in (\ref{vinet:eq4.10}). Observing further that in terms of $t$, $\psi_{0,0}$ given in (\ref{vinet:eq4.5}) reads
    \begin{eqnarray} \label{vinet:eq7.46}   
\psi_{0,0} = \frac{1}{(\ch \rho)^{1/2}} \ee^{-\sigma^2(1+t)/2}
   \end{eqnarray}      
  and that
   \begin{eqnarray} \label{vinet:eq7.47}   
  \ee^{-ik\delta} \left( \frac{t}{2}\right)^{k/2} = \bigg( \frac12 \ee^{-i\theta} \Th \rho\bigg)^{k/2},
     \end{eqnarray}      
  we see that     (\ref{vinet:eq7.44}) coincides with the expression for $\psi_{0,k}$ given in  (\ref{vinet:eq4.11}).
  
  Let us now look at the bottom component of  (\ref{vinet:eq7.40}). Using (\ref{vinet:eqA.13}) we have in this case that
  \begin{eqnarray} \label{vinet:eq7.48}     
  \psi_{1,k} = D_k + E_k
      \end{eqnarray}         
where      
 \numparts \label{vinet:eq7.49}   
\begin{eqnarray} \label{vinet:eq7.49a}  
\fl  D_k =  \frac{1}{\sqrt{k!}} (-\sigma \ee^{-i\delta})^k  \ee^{-\sigma^{2/2}} \cdot \frac{\sigma \ee^{i\delta}}{(\ch \rho)^{3/2}}\nonumber\\[2\jot]
 \cdot \ee^{-\sigma^2 t/2} \sum_{\ell=0}^{[k/2]} \frac{(-k)_{2\ell}}{(1+t)^{2\ell -k}} \cdot \frac{1}{\ell !} \cdot \bigg( -\frac{t}{2\sigma^2}\bigg)^{\ell}  
        \end{eqnarray}           
 and
 \begin{eqnarray} \label{vinet:eq7.49b}  
\fl   E_k = \frac{1}{\sqrt{k!}} (-\sigma \ee^{-i\delta})^k  \ee^{-\sigma^{2/2}} \cdot \frac{\sigma \ee^{i\delta}}{(\ch \rho)^{3/2}}\nonumber\\[2\jot]
 \cdot \ee^{-\sigma^2 t/2} \sum_{\ell=0}^{[k/2]} \frac{(-k)_{2\ell}(2\ell-k)}{\sigma^2(1+t)^{2\ell -k+1}} \cdot \frac{1}{\ell !} \cdot \bigg( -\frac{t}{2\sigma^2}\bigg)^{\ell},
        \end{eqnarray}               
\endnumparts
correspond to the two terms occurring in the sum on the r.h.s.  of  (\ref{vinet:eqA.13}). Comparing  (\ref{vinet:eq7.49a}) with  (\ref{vinet:eq7.41}) we see that
   \begin{eqnarray} \label{vinet:eq7.50}  
D_k = \frac{\sigma \ee^{i\delta}}{\ch \rho} \psi_{0,k}
        \end{eqnarray}            
and thus we readily have
   \begin{eqnarray} \label{vinet:eq7.51}  
D_n = \frac{1}{\sqrt{k!}} \frac{\sigma}{(\ch \rho)^{3/2}} \ee^{-i(k-1)\delta} \bigg(\frac{t}{2} \bigg)^{k/2} \ee^{-\sigma^2 (1+t)/2} H_k(s).
  \end{eqnarray}        
As for $E_k$, with
   \begin{eqnarray} \label{vinet:eq7.52}  
(-k)_{2\ell} (2\ell - k) = (-1)^{2\ell + 1} \frac{k!}{(k-2\ell -1)!}
  \end{eqnarray}    
  and using this time the factorization
    \begin{eqnarray} \label{vinet:eq7.53}   
  \bigg( \frac{t}{2\sigma^2}\bigg)^{\ell} = \bigg( \frac{\sqrt{2} \sigma}{t^{1/2}}\bigg)^{k-2\ell-1} \cdot \bigg(\frac{t}{2\sigma^2}\bigg)^{(k-1)/2}
   \end{eqnarray}    
   one arrives at
  \begin{eqnarray} \label{vinet:eq7.54}      
\fl   E_k =  \frac{1}{\sqrt{k!}} (\sigma \ee^{-i\delta})^k \ee^{- \sigma^2(1+t)/2} \cdot \frac{\ee^{i\delta}}{\sigma(\ch \rho)^{3/2}} \cdot \bigg(\frac{t}{2\sigma^2}\bigg)^{(k-1)/2}\nonumber\\[2\jot]
    \cdot k \left[ \sum_{\ell=0}^{[(k-1)/2]} \frac{(-1)^{\ell}}{\ell ! (k-2\ell -1)!} \bigg(-\frac{2}{\sqrt{2}} \sigma (t^{1/2} + t^{-1/2})\bigg)\right].
   \end{eqnarray}      
   As before the term in brackets is identified as $H_{k-1}(s)$ with the help of (\ref{vinet:eqA.3}), so that
       \begin{eqnarray} \label{vinet:eq7.55}    
   E_k = \frac{k}{\sqrt{k!}} \frac{1}{(\ch \rho)^{3/2}} \ee^{-{\sigma^2(1+t)}/{2}}\ee^{-i(k-1)\delta} \bigg( \frac{t}{2}\bigg)^{(k-1)/2} H_{k-1}(s).
   \end{eqnarray}        
Combining (\ref{vinet:eq7.51}) and (\ref{vinet:eq7.55}), we find
       \begin{eqnarray} \label{vinet:eq7.56}    
\psi_{1,k} &= D_k + E_k\nonumber\\
 &= \frac{1}{\sqrt{k!}} \frac{1}{\ch \rho} \bigg( \ee^{-2i\delta} \frac{t}{2}\bigg)^{k/2} \cdot \ee^{i\delta} \cdot \sqrt{2} t^{-1/2}\nonumber\\[2\jot]
&\quad \cdot \left[ \sigma\bigg(\frac{t}{2}\bigg)^{1/2} H_k(s) + k H_{k-1}(s) \right] \psi_{0,0}.
   \end{eqnarray}        
In order to recover (\ref{vinet:eq4.15}), we use the recurrence relation  (\ref{vinet:eqA.1}) to replace the $H_{k-1}(s)$ by
       \begin{eqnarray} \label{vinet:eq7.57}    
k H_{k-1}(s) = sH_k(s) -\frac12 H_{k+1}(s).
   \end{eqnarray}    
In view of (\ref{vinet:eq7.45}),
   \begin{eqnarray} \label{vinet:eq7.58}    
\fl      \left[ \frac{\sigma}{\sqrt{2}} t^{1/2} H_k(s) + k H_{k-1}(s) \right]\nonumber \\
 = -\frac{\sigma}{\sqrt{2}} t^{-1/2} H_k(s) - \frac12 H_{k+1}(s)\nonumber\\[2\jot]
 =- \frac{t^{-1/2}}{\sqrt{2}}\bigg( \sigma H_k(s) + \bigg(\frac{t}{2}\bigg)^{1/2} H_{k+1}(s)\bigg)
   \end{eqnarray}    
from where it follows that
  \begin{eqnarray} \label{vinet:eq7.59}    
\fl    \psi_{1,k}  =  -\frac{1}{\sqrt{k!}} \cdot \frac{1}{\sh \rho}   \bigg(\ee^{-2i\delta} \frac{t}{2}\bigg)^{k/2}\nonumber\\[2\jot]
 \cdot \ee^{i(\theta -\delta)} \cdot \bigg\{ \sigma H_k(s) + \bigg( \frac{t}{2}\bigg)^{1/2} H_{k+1}(s) \bigg\} \psi_{0,0}
\end{eqnarray}        
in accordance with  (\ref{vinet:eq4.15}). Let us record that from (\ref{vinet:eq7.56}), using $H_{k-1}(s)$, we have alternatively to (\ref{vinet:eq4.15}):
  \begin{eqnarray} \label{vinet:eq7.60} 
\fl\psi_{1,k} =  \frac{1}{\sqrt{k!}} \cdot \frac{1}{\ch \rho} \bigg( \ee^{-i\theta} \frac{\Th \rho}{2} \bigg)^{k/2}\nonumber\\
  \cdot \left[ \sigma \ee^{i\delta} H_k(s) + \sqrt{2} \frac{k \ee^{i\theta/2}}{\Th^{1/2} \rho} H_{k-1}(s)\right] \psi_{0,0}.   
\end{eqnarray}         

\section{Generating functions and 2-variable Hermite polynomials} \label{vinet:section8}

We shall discuss in this section, generating functions for the $\Sch_1$ group representation elements. We shall examine in turn a 2-variable and a 1-variable generating function. As we shall see below, in the former case, 2-variable Hermite polynomials will appear and hence be shown to have connections with the special functions that have arisen so far.

\subsection{2-variable generating function}\label{vinet:subsection8.1}
      
Let $x$ and $y$ be two complex indeterminates. Following Perelomov and Popov~\cite{vinet:ref11}, we define the 2-variable generating function $G(x,y)$ as follows:
 \begin{eqnarray} \label{vinet:eq8.1} 
G(x,y) &= \sum_{k,n=0}^{\infty} \frac{\bar{x}^k}{\sqrt{k!}} \frac{y^n}{\sqrt{n!}} \, \psi_{n,k}\nonumber\\
&= \sum_{k,n=0}^{\infty} \frac{\bar{x}^k}{\sqrt{k!}} \frac{y^n}{\sqrt{n!}} \langle k \vert S \vert n\rangle.
\end{eqnarray}     
 Clearly, $G(x,y)$ can be looked at as the matrix element of $S$ between the coherent states $\vert x \rangle$ and $\vert y\rangle$ :
  \begin{eqnarray} \label{vinet:eq8.2} 
 G(x,y) = \langle x \vert S \vert y \rangle.
\end{eqnarray}    
We shall rely on the properties (\ref{vinet:eq1.11a}, \ref{vinet:eq1.11b}, \ref{vinet:eq1.11c}) of these special states, to first evaluate $G(x,y)$. Once again we shall use the disentangled form of $S$ as provided by formulas (\ref{vinet:eq4.2}) and (\ref{vinet:eq4.4}). It is straightforward to see that
  \begin{eqnarray} \label{vinet:eq8.3} 
\langle x \vert \ee^{va - \bar{v} a^+} 
&= \langle x \vert \ee^{- v \bar{v}/2} \ee^{-\bar{v}a^+} \ee^{va}\nonumber\\
&= \ee^{-v\bar{v}/2} \ee^{-\bar{v}\bar{x}} \langle x + \bar{v}\vert.
\end{eqnarray}    
We have thus
  \begin{eqnarray} \label{vinet:eq8.4} 
\fl    \langle x\vert S \vert y \rangle =  \ee^{- v \bar{v}/2} \ee^{-\bar{v} \bar{x}} 
\langle x + \bar{v} \vert \exp \bigg[ -\frac12 \ee^{-i\theta} \Th \rho\, (a^+)^2\bigg]\nonumber\\[2\jot]
 \cdot \exp \bigg[ -\ln (\ch \rho)\bigg(a^+a+\frac12\bigg)\bigg] \cdot \exp \bigg[ \frac12 \ee^{i\theta} \Th\rho\, a^2 \bigg] \vert y \rangle.
\end{eqnarray}  
 As an intermediary step in the determination of the r.h.s. of (\ref{vinet:eq8.4}) , one needs :
  \begin{eqnarray} \label{vinet:eq8.5}  
\fl      \langle x + \bar{v} \vert \exp \bigg[ -(\ln \ch \rho) \bigg(a^+a + \frac12\bigg)\bigg]  \vert y\rangle\nonumber\\
  = \sum_{\ell, m=0}^{\infty} \langle x + \bar{v} \vert \ell \rangle \, \langle \ell \vert \exp \bigg[ - (\ln \ch \rho)\bigg( a^+a + \frac12\bigg)\bigg] \vert m \rangle \, \langle m \vert y \rangle\nonumber\\
 = \sum_{\ell,m=0}^{\infty} \frac{(\bar{x} + v)^{\ell}}{\sqrt{\ell!}} \exp \bigg[ - \bigg(m + \frac12\bigg) \ln (\ch \rho)\bigg] \delta_{\ell m} \frac{y^m}{\sqrt{m!}}\nonumber\\
 = (1 - \Th^2 \rho)^{1/4} \sum_{m=0}^{\infty} \frac{1}{m!} [(1-\Th^2\rho)^{1/2}(\bar{x} + v)y]^m\nonumber\\
 = (1-\Th^2 \rho)^{1/4} \exp[(1-\Th^2\rho)^{1/2} (\bar{x}+v)y],
\end{eqnarray}            
 where we used (\ref{vinet:eq1.11c}) and employed the identity $\ch \rho = (1-\Th^2\rho)^{-1/2}$.
 
 With the help of (\ref{vinet:eq1.11b}), it is easy to complete the calculation to find that
  \begin{eqnarray} \label{vinet:eq8.6}   
\fl     G(x,y) 
  =  (1-\Th^2\rho)^{1/4} \cdot \exp \bigg( -\frac12 v \bar{v} - \frac12 \ee^{-i\theta} \Th \rho\, v^2\bigg)\nonumber\\
 \cdot \exp \bigg\{ - \frac12 \ee^{-i\theta} \Th \rho\, \bar{x}^2 + (1-\Th^2\rho)^{1/2} \bar{x}y\nonumber\\
  + \frac12 \ee^{i\theta} \Th\rho\, y^2 + (- \bar{v} - \ee^{-i\theta}\Th\rho\, v)\bar x + (1-\Th^2\rho)^{1/2} vy \bigg\}.
\end{eqnarray}            
Now let
  \begin{eqnarray} \label{vinet:eq8.7} 
z_1 = \bar{x}; \qquad z_2 = y.
\end{eqnarray}                
 Given that the exponent in the last term of   (\ref{vinet:eq8.6}) is quadratic in the indeterminates, we can rewrite $G(x,y)$ in the form
   \begin{eqnarray} \label{vinet:eq8.8} 
\fl G(x,y) =  (1-\Th^2\rho)^{1/4} \cdot \exp \bigg(-\frac12 v \bar{v} - \frac12 \ee^{-i\theta} \Th\rho\, v^2\bigg)\nonumber\\
  \cdot \exp \left\{ \sum_{i,j=1,2} a_{ij}(c_iz_j - z_iz_j/2) \right\}
  \end{eqnarray}            
where the $2 \times 2$ symmetric matrix $[a_{ij}]$ has for its elements:
   \begin{eqnarray} \label{vinet:eq8.9} 
a_{11} &= \ee^{-i\theta}\Th\rho,   \qquad a_{22} = -\ee^{i\theta}\Th \rho,\nonumber\\
a_{12} &= a_{21} = -(1-\Th^2\rho)^{1/2},
  \end{eqnarray}          
and where
 \numparts
   \begin{eqnarray} \label{vinet:eq8.10} 
c_1 &= -v(1 + \ee^{i\theta} \Th \rho), \label{vinet:eq8.10a} \\
c_2 &= \frac{v}{\ch \rho}.\label{vinet:eq8.10b} 
  \end{eqnarray}       
\endnumparts
Consulting formula  (\ref{vinet:eqA.18}), we conclude that $G(x,y)$ is up to a factor, the generating function of the 2-variable Hermite polynomials $H_{k,n}(c_1,c_2)$. We found indeed that
  \begin{eqnarray} \label{vinet:eq8.11} 
   G(x,y)  &= \sum_{k,n=0}^{\infty} \frac{\bar{x}^k}{\sqrt{k!}} \frac{y^n}{\sqrt{n!}} \, \psi_{n,k}\nonumber\\
&= (\ch \rho)^{-1/2} \cdot \exp \bigg[ -\frac12 \sigma^2 (1 + \ee^{i(2\delta - \theta)}\Th \rho) \bigg]\nonumber\\
&\quad  \cdot \sum_{k,n=0}^{\infty} \frac{\bar{x}^k}{k!} \frac{y^n}{n!} H_{k,n} (c_1,c_2)  
  \end{eqnarray}           
from where one immediately reads that
   \begin{eqnarray} \label{vinet:eq8.12} 
\psi_{n,k} = \frac{1}{\sqrt{k!n!}} H_{k,n} \bigg[ -\sigma \ee^{i\delta} (1+ \ee^{i\theta} \Th\rho), \frac{\sigma \ee^{i\delta}}{\ch \rho}\bigg] \psi_{0,0}
  \end{eqnarray}       
  recalling  (\ref{vinet:eq4.5}) and putting $v = \sigma \ee^{i\delta}$. In other words, the $\Sch_1$ representation matrix elements $\psi_{n,k}$ are essentially 2-variable Hermite polynomials. As a consequence the relation $\Psi_{n,k} = \mathcal{P}_n(k)\Psi_{0,k}$ (in view of  (\ref{vinet:eq4.11}) and  (\ref{vinet:eq4.15}) provides a relation between 2- and 1- variable Hermite polynomials. It reads
 \begin{eqnarray} \label{vinet:eq8.13} 
\fl  \left(\begin{array}{l}
 \frac{1}{\sqrt{2n!}} H_{k,2n}(c_1,c_2)\\[2\jot]
\frac{1}{\sqrt{(2n+1)!}} H_{k,2n+1}(c_1,c_2)
  \end{array}
  \right) =
  \bigg( \frac12 \ee^{-i\theta} \Th\rho\bigg)^{k/2} \nonumber\\[2\jot]
 \cdot \mathcal{P}_n(k)
   \left(\begin{array}{l}
 H_k(s)\\[2\jot]
-\frac{1}{\sh \rho}\bigg[ \sigma \ee^{i(\theta-\delta)}H_k(s)  \\
  \qquad\qquad\qquad + \bigg( \frac12 \ee^{i\theta}\Th\rho\bigg)^{1/2}H_{k+1}(s)\bigg] 
  \end{array}
  \right)     
  \end{eqnarray}         
  with $c_1$, $c_2$ and $s$, respectively given by (\ref{vinet:eq8.10a})--(\ref{vinet:eq8.10b}) and (\ref{vinet:eq4.10}). Analogously, (\ref{vinet:eq7.38}) entails an expansion of the 2-variable Hermite polynomials $H_{k,n}(c_1,c_2)$ in terms of Charlier and Meixner polynomials.
  
  \subsection{1-variable generating function} \label{subsection:vinet8.2}
  
  It is also instructive to consider the 1-variable generating function $F_k(y)$ defined by
 \begin{eqnarray} \label{vinet:eq8.14} 
  F_k(y) = \sum_{n=0}^{\infty} \frac{y^n}{\sqrt{n!}} \, \psi_{n,k} = \sum_n \frac{y^n}{\sqrt{n!}} \langle k \vert S \vert n\rangle.
   \end{eqnarray}  

It is obviously related as follows to the 2-variable generating function $G(x,y)$:
\begin{eqnarray} \label{vinet:eq8.15}
G(x,y) = \sum_{k=0}^{\infty} \frac{\bar{x}^k}{\sqrt{k!}} F_k(y)
  \end{eqnarray}
which can thus be looked at also, as the generating function for the $F_k (y)$. Let us return to expression (\ref{vinet:eq8.6}) for $G(x,y)$, this time, treating $y$ as a parameter and keeping only $x$ as a variable. Take
\begin{eqnarray} \label{vinet:eq8.16}
z = \bigg( \frac12 \ee^{-i\theta} \Th \rho\bigg)^{1/2} \bar{x}
  \end{eqnarray}
and
\begin{eqnarray} \label{vinet:eq8.17}
r =
\left\{ \frac{\ee^{i\theta/2}}{\sqrt{\sh 2\, \rho}} y - \frac{\sigma}{\sqrt{2}} [ \ee^{i(\delta - \theta/2)} \Th^{1/2}\rho  + \ee^{-i(\delta - \theta/2)} \Th^{-1/2} \rho  ] \right\}.
  \end{eqnarray}
It is a simple matter to check that $G(x,y) = \widetilde{G}(z,y)$ can be written as follows:
 \begin{eqnarray}
\fl \widetilde{G}(z,y) = \exp \bigg[ (1-\Th^2\rho)^{1/2} \sigma \ee^{i\delta} y + \frac12 \ee^{i
\theta} \Th\rho\, y^2\bigg] \nonumber \\
\cdot \exp(2rz - z^2) \cdot \psi_{0,0}. \label{vinet:eq8.18}
  \end{eqnarray}
With the help of (\ref{vinet:eqA.2}), we recognize in (\ref{vinet:eq8.18}) the presence of the generating function $\exp(2rz-z^2)$ of the 1-variable Hermite polynomials. We therefore have
\begin{eqnarray} \label{vinet:eq8.19}
\fl G(x,y)=  \exp \bigg[ \frac{\sigma \ee^{i\delta}}{\ch \rho} y + \frac12 \ee^{i\theta} \Th\rho\, y^2\bigg] \nonumber\\[2\jot]
  \cdot \left( \sum_{k=0}^{\infty} \frac{1}{k!} \bigg( \frac12 \ee^{-i\theta}\Th\rho\bigg)^{k/2} H_k (r) \bar{x}^k \right) \cdot \psi_{0,0}
  \end{eqnarray}
from where we find, given (\ref{vinet:eq8.15}), the following 1-variable generating function for the $\Sch_1$ representation matrix elements
\begin{eqnarray} \label{vinet:eq8.20}
\fl F_k(y) 
 = \sum_{n=0}^{\infty} \frac{y^n}{\sqrt{n!}} \, \psi_{n,k} =\frac{1}{\sqrt{k!}} \bigg( \frac{1}{2}  \ee^{-i\theta}\Th \rho\bigg)^{k/2}\nonumber\\
  \cdot \exp \bigg[ \frac{\sigma\ee^{i\delta}}{\ch \rho} y + \frac12 \ee^{i\theta} \Th\rho\, y^2 \bigg] \cdot H_k(r) \cdot \psi_{0,0}.
   \end{eqnarray}           
 
 Interestingly, it is expressed in terms of 1-variable Hermite polynomials in the variable $r$ of  (\ref{vinet:eq8.17}), which can also be written
  \begin{eqnarray} \label{vinet:eq8.21}
 r = \frac{\ee^{-i\theta/2}}{(\sh 2\, \rho)^{1/2}} y + s
    \end{eqnarray}  
 where $s$, as defined in (\ref{vinet:eq4.10}), is the argument of the Hermite polynomials in the expressions of $\psi_{0,k}$ and $\psi_{1,k}$ (see Section~\ref{vinet:section4}). We can now have a last additional look at these two fundamental matrix elements in relation with the generating function.
 
 Note that
 \begin{eqnarray} \label{vinet:eq8.22}
 F_k(0) = \psi_{0,k}.
   \end{eqnarray}
   
Evaluating (\ref{vinet:eq8.20}) at $y=0$, gives
 \[
\psi_{0,k} = \frac{1}{\sqrt{k!}} \bigg( \frac12 \ee^{-i\theta} \Th\rho \bigg)^{k/2} H_k(s) \, \psi_{0,0},  
\]
   precisely the expression   (\ref{vinet:eq4.11})  obtained in Section~\ref{vinet:section4}.
   
   Similarly
  \begin{eqnarray} \label{vinet:eq8.23}  
   \psi_{1,k}  &= \frac{\dd}{\dd y} F(y) \bigg|_y=0\nonumber\\[2\jot]
    &= \frac{1}{\sqrt{k!}} \bigg( \frac{1}{2} \ee^{i\theta} \Th\rho\bigg)^{k/2}\nonumber\\[2\jot]
   &\quad \cdot \left[ \frac{\sigma \ee^{i\delta}}{\ch\rho} H_k(r) +
  \frac{\ee^{i\theta/2}}{\sqrt{2\sh\rho \ch \rho}} \frac{\dd}{\dd r} H_k(r)\right]_{y=0}
  \end{eqnarray}
which is found to coincide with  (\ref{vinet:eq7.60}), using the Appell property  (\ref{vinet:eqA.4}) of the Hermite polynomials.

\section{Position representation and affine transformations of Hermite polynomials}\label{vinet:section9}

The matrix elements $\psi_{n,k}$ of the $\Sch_1$ representation have been obtained quite generally from the structure of the group and in particular through an embedding of its generators in the Heisenberg (envelopping) algebra. When particular realizations are brought in, typically, more special functions become associated to the symmetry group and further relations with the special functions already present in the general matrix elements often result~\cite{vinet:ref16a}. We provide a brief illustration of that in this section by showing how formulas for the affine transformations of Hermite polynomials can be obtained in this fashion.

In the position representation, the creation and annihilation operators take the familiar form  (\ref{vinet:eq1.6})  and the vector states are realized by the wave functions $\Phi_n(x) = \langle x \vert n \rangle$ given in (\ref{vinet:eq1.10}), where Hermite polynomials intervene\footnote{In this section $\vert x\rangle$ is the eigenstate with real eigenvalue $x$ of the position operator $(a+a^+)/\sqrt{2}$ and should not be confused with a coherent state.}. Upon exponentiating the (generalized) vector fields arising in this picture, the Schr\"odinger group $\Sch_1$ is represented projectively as a transformation group on functions. Denote by $\Phi_n^S(x)$ the transform of $\Phi_n(x)$ under $S$ which we can write
\begin{eqnarray} \label{vinet:eq9.1}
\Phi_n^S (x) = \langle x \vert S\vert n \rangle.
  \end{eqnarray}
Parallely, it can be equated to
\begin{eqnarray} \label{vinet:eq9.2}
\langle x \vert S \vert n\rangle 
&= \sum_{k=0}^{\infty} \langle x \vert k \rangle \, \langle x \vert S \vert n \rangle\nonumber\\
&= \sum_{k=0}^{\infty} \, \psi_{n,k} \Phi_k(x)
  \end{eqnarray}            
and therein lie identities for the Hermite polynomials that generalize somehow their generating function relation as well as that of the Laguerre polynomials (see below).

We shall take $v$ and $w$ real, thereby focussing on affine transformations. When $\delta = \theta = 0$,
\begin{eqnarray} \label{vinet:eq9.3}
S = \ee^{\sigma(a-a^+)} \ee^{\rho(a^2 - (a^+)^2)/2}.
  \end{eqnarray}   
We already noted in (\ref{vinet:eq1.12})--(\ref{vinet:eq1.13}) that $a-a^+ = \sqrt{2}\, \DD/\dd x$ and $a^2-(a^+)^2 = 2x\, \DD/\dd x + 1$. In this case, $S$ thus consists of dilations with parameter $\rho$, followed by translations with parameter $\sqrt{2}\, \sigma$. Standard Lie theory gives
\begin{eqnarray} \label{vinet:eq9.4}
\langle x \vert S \vert n \rangle 
&= \ee^{\sqrt{2}\, \sigma {\dd}/{\dd x}} \cdot \ee^{\rho({\dd}/{\dd x} + 1/2)} \Phi_n(x)\nonumber\\[2\jot]
&= \ee^{\rho/2} \Phi_n[\ee^{\rho}(x + \sqrt{2}\, \sigma)].
  \end{eqnarray}   
Let us first consider separately the cases $\rho = 0$ and $\sigma = 0$. When $\rho = 0$, from 
(\ref{vinet:eq9.2}), we have
\begin{eqnarray} \label{vinet:eq9.5}
\Phi_n(x + \sqrt{2}\, \sigma) = \sum_{k=0}^{\infty} \chi_{n,k} \Phi_k(x)
  \end{eqnarray}   
where $\chi_{n,k}$ are the specializations of $\psi_{n,k}$ to $\rho = 0$ given in (\ref{vinet:eq7.13}). Using this formula in (\ref{vinet:eq9.5}), together with the precise form (\ref{vinet:eq1.10}) of $\Phi_n$ yields 
\begin{eqnarray} \label{vinet:eq9.6}
H_n(x + \sqrt{2}\, \sigma)
 = (\sqrt{2}\, \sigma)^n \ee^{(\sqrt{2}\, \sigma x + \sigma^2/2)}  \cdot \sum_{k=0}^{\infty} \frac{(-\sigma)^k}{2^{k/2}k!} C_n(k; \sigma^2)H_k(x)
  \end{eqnarray}   
which provides an expression for the transform of Hermite polynomials under translations~(\cite{vinet:ref16a}, \cite{vinet:ref15}). Note that (\ref{vinet:eq9.6}) can be viewed as a generalization of the generating function identity (\ref{vinet:eqA.2}) which is recovered from (\ref{vinet:eq9.6}) when $n=0$ since $H_0(x) = 1 = C_0(k;\sigma^2)$.

When $\sigma = 0$, we recall from Section~\ref{vinet:section7} that $\psi_{n,k} = \varphi_{n,k}$ and that only $\varphi_{2n,2k}$ and $\varphi_{2n+1,2k+1}$ are non-zero. Eqs.~(\ref{vinet:eq9.4}) and (\ref{vinet:eq9.2}) therefore amount to the pair of equations:
 \numparts
\begin{eqnarray}\label{vinet:eq9.7}
\ee^{\rho/2} \Phi_{2n} (\ee^{\rho}x)
&= \sum_{k=0}^{\infty} \varphi_{2n,2k} \Phi_{2k}(x) \label{vinet:eq9.7a}\\
\ee^{\rho/2} \Phi_{2n+1} (\ee^{\rho}x)
&= \sum_{k=0}^{\infty} \varphi_{2n+1,2k+1} \Phi_{2k+1}(x).  \label{vinet:eq9.7b}
 \end{eqnarray}
\endnumparts
 Again, substituting for $\Phi_n(x)$ and using the explicit forms of the matrix elements $\varphi_{n,k}$ given in  (\ref{vinet:eq7.3}), we find from (\ref{vinet:eq9.7a}, \ref{vinet:eq9.7b}) after some simplifications, the following two formulas
 \numparts\label{vinet:eq9.8}
 \begin{eqnarray}
\fl H_{2n}(\ee^{\rho}x)
  =  \frac{(2n)!}{n!} \ee^{-\rho/2} \ee^{(\ee^{2\rho}-1)x^2/2} \frac{ \Th^n\rho }{\ch^{1/2} \rho}\nonumber\\
  \cdot \sum_{k=0}^{\infty} \frac{(-1)^k}{2^{2k}k!} \Th^k \rho M_n\bigg( k; \frac12, \Th^2\rho\bigg) H_{2k}(x) \label{vinet:eq9.8a}
   \end{eqnarray}
   
  \begin{eqnarray} \label{vinet:eq9.8b}
 \fl H_{2n+1}(\ee^{\rho}x)
  =  \frac{(2n+1)!}{n!} \ee^{-\rho/2} \ee^{(\ee^{2\rho}-1)x^2/2} \frac{\Th^n\rho}{ \ch^{3/2} \rho }\nonumber\\
  \cdot \sum_{k=0}^{\infty} \frac{(-1)^k}{2^{2k}k!} \Th^k \rho M_n\bigg( k; \frac32, \Th^2\rho\bigg) H_{2k+1}(x)
   \end{eqnarray} 
\endnumparts
   which offer expressions for the transforms of even and odd Hermite polynomials under dilations of their argument. In this case, equations (\ref{vinet:eq9.8a}, \ref{vinet:eq9.8b}) can be looked at as generalizations of the generating functions relation for the Laguerre polynomials, keeping in mind the connection  (\ref{vinet:eqA.5a}--\ref{vinet:eqA.5b}) between the Hermite and Laguerre families. Indeed, when $n=0$, it is straightforward to check that  (9.8) reduces to  (\ref{vinet:eqA.7}).
   
   Formulas for the transform of $H_n(x)$ under the general affine coordinate transformation
 \begin{eqnarray} \label{vinet:eq9.9}   
   x' = \ee^{\rho} (x + \sqrt{2}\, \rho)
    \end{eqnarray}   
are obtained by compounding  (\ref{vinet:eq9.6}) and  (\ref{vinet:eq9.8}), which is in keeping with the convolution  (\ref{vinet:eq7.38}) found in Section~\ref{vinet:section7} for the general matrix elements. While this is straightforward to achieve, the resulting expressions are a bit cumbersome and will not be recorded here. Let us rather point out that in terms of the matrix polynomials $\mathcal{P}_n(k)$, we have
\begin{eqnarray} \label{vinet:eq9.10}
\fl\left(\begin{array}{l l}
  \frac{1}{2^n\sqrt{(2n)!}} H_{2n} (x')\\[2\jot]
   \frac{1}{2^{n+1/2}\sqrt{(2n+1)!}} H_{2n+1} (x')
  \end{array}
  \right) =
  \ee^{(x'^2 - x^2)/2-  {\rho}/{2}} \sum_{k=0}^{\infty} \frac{H_k(x)}{2^{k/2}\sqrt{k!}} \mathcal{P}_n(k) \Psi_{0,k}
     \end{eqnarray}       
 where $x'$ is as in  (\ref{vinet:eq9.5}). This equation follows from the transcription of (\ref{vinet:eq9.2}) in vector notation
 \begin{eqnarray} \label{vinet:eq9.11}
 \left(\begin{array}{l l}
  \Phi_{2n}^S(x)\\[2\jot]
  \Phi_{2n+1}^S(x)
  \end{array}
  \right) = \sum_{k=0}^{\infty} \, \Psi_{n,k} \Phi_k(x)
   \end{eqnarray}
   and from using our fundamental result, namely that $\Psi_{n,k} = \mathcal{P}_n(k)\Phi_k(x)$. The reader will remember, as shown throughout the paper, that the MOPs $\mathcal{P}_n(k)$ can be obtained in a variety of ways: recurrence relation, difference equations, Rodrigues' formula etc.
   
   \section{Concluding remarks} \label{vinet:section10}
   In summary, this paper has been concerned with the Schr\"odinger group $\Sch_1$ in one-dimension---a fundamental symmetry group of non-relativistic physics. Basically, we have achieved an explicit and thorough determination of the representations of that group on the natural Heisenberg algebra module. We have found that the matrix elements of this representation are expressed in terms of matrix orthogonal polynomials.
   
   Our study has provided in fact, what is probably the most simple example of matrix orthogonal polynomials in a physical context. In view of the efforts currently deployed to develop the theory of matrix orthogonal polynomials, we trust that this example will provide insightful illustration and offer a useful test bed.
   
   Of particular interest is also that we could present a full symmetry analysis; here again a rare case in dealing with MOPs. This has allowed to realize a systematic and exhaustive characterization (3-term recurrence relation, orthogonality relations, difference equation, ladder operators, Rodrigues' formula, generating functions and scalar contents).
   
   Let us mention in closing possible directions for further study. The 2-mode extension has an obvious physical interest. Examining the possible supersymmetric extension, $q$- deformation or parabosonic version would certainly be worthwhile. Finally, it would generally be desirable to explore other physical and algebraic contexts where matrix orthogonal polynomials will arise.
   
   \ack
   The authors are thankful to A. Gr\"unbaum, W. Miller Jr., S. Tsujimoto, and P.~Winternitz for useful discussions. The work of L.V. is supported in part through funds provided by the Natural Science and Engineering Research Council (NSERC) of Canada. A.Z. is grateful for the hospitality extended to him at the Centre de recherches math\'ematiques (Montr\'eal).
   
   \appendix
   \section{A compendium of formulas for orthogonal polynomials} \label{vinet:appendixA.1} 
   
   In order to make the paper self-contained and to specify our notations and conventions, we collect in this Appendix most formulas for orthogonal polynomials that are used in the body of the article.
  
 \subsection{Hermite polynomials~\cite{vinet:ref16}} \label{vinet:subsectionA.1} 
 The Hermite polynomials $H_n(x)$ can be defined by the following 3-terms recurrence relations:
\begin{eqnarray} \label{vinet:eqA.1}     
H_{n+1}(x) + 2nH_{n-1}(x) - 2xH_n(x) = 0
\end{eqnarray}     
with $H_0(x) = 1$. They have the generating function
\begin{eqnarray} \label{vinet:eqA.2}  
\sum_{n=0}^{\infty} H_n(x) \frac{z^n}{n!} = \exp (2x z-z^2),
    \end{eqnarray}           
 and the explicit expression
 \begin{eqnarray} \label{vinet:eqA.3} 
 H_n(x) = n! \sum_{m=0}^{[n/2]} \frac{(-1)^m(2x)^{n-2m}}{m!(n-2n)!},
   \end{eqnarray}
   where $[s]$ denotes the integer part of $s$. 
   
 They enjoy the Appell property
\begin{eqnarray} \label{vinet:eqA.4} 
 \frac{d}{\dd x} H_n(x) = 2n H_{n-1}(x).
   \end{eqnarray}
 The Hermite polynomials can be expressed in terms of Laguerre polynomials $L^{\alpha}_{n}(x)$ (see below):
 \numparts
 \begin{eqnarray} \label{vinet:eqA.5} 
 H_{2n}(x) &= (-1)^n 2^{2n} n! L_n^{-1/2}(x^2),\label{vinet:eqA.5a}  \\[2\jot]
 H_{2n+1}(x) &= (-1)^n 2^{2n+1} n! x L_n^{1/2}(x^2).\label{vinet:eqA.5b} 
   \end{eqnarray}
\endnumparts

  \subsection{Laguerre polynomials~\cite{vinet:ref16}} \label{vinet:subsectionA.2} 
  
  The Laguerre polynomials $L_n^{\alpha}(x)$, $\alpha > -1$ obey the 3-term recurrence relation
\begin{eqnarray} \label{vinet:eqA.6}   
  (n+1) L_{n+1}^{\alpha}(x) - (2n + \alpha + 1 -x) L_n^{\alpha}(x) + (n+\alpha) L_{n-1}^{\alpha}(x) = 0
   \end{eqnarray}  
  and have the generating function  
\begin{eqnarray} \label{vinet:eqA.7}     
  \sum_{n=0}^{\infty} L_n^{\alpha}(x) z^n = (1-z)^{-\alpha - 1} \exp \frac{xz}{z-1}.
    \end{eqnarray}   
  
  \subsection{Charlier polynomials~\cite{vinet:ref17}}     \label{vinet:subsectionA.3}  
 The Charlier polynomials $c_n(x;a)$, $a > 0$ are polynomials of a discrete variable and satisfy the 3-term recurrence relation      
 \begin{eqnarray} \label{vinet:eqA.8}
 -x c_n(x;a) = ac_{n+1} (x;a) - (n+a)c_n(x;a) + nc_{n-1}(x;a).
     \end{eqnarray}   
 We shall also record the normalized recurrence relation obeyed by the monic version $p_n(x) = x^n + O(n-1)$ of the polynomials. Here
 \begin{eqnarray} \label{vinet:eqA.9}
 xp_n(x) = p_{n+1}(x) + (n+a)p_n(x) + nap_{n-1}(x)
     \end{eqnarray}   
 where 
 \begin{eqnarray} \label{vinet:eqA.10}
 c_n(x;a) = \bigg( -\frac{1}{a}\bigg)^n p_n(x).
    \end{eqnarray}      
    A generating formula for the Charlier polynomials is
 \begin{eqnarray} \label{vinet:eqA.11}   
    \sum_{n=0}^{\infty} c_n(x;a) \frac{t^n}{n!} = \ee^t \bigg( 1-\frac{t}{a}\bigg)^x.
     \end{eqnarray}       
    Using umbral calculus, Gessel~\cite{vinet:ref14} has obtained a formula for a sum involving only the polynomials of even degree. In our conventions it reads:
\begin{eqnarray} \label{vinet:eqA.12}    
    \sum_{n=0}^{\infty} c_{2n}(x;a) \frac{1}{n!} \bigg( -\frac{1}{2}at \bigg)^n = \ee^{- at/2} \sum_{\ell=0}^{\infty} \frac{(-x)_{2\ell}}{(1+t)^{2\ell -x}} \frac{1}{\ell !} \bigg( - \frac{t}{2a}\bigg)^{\ell}.
     \end{eqnarray}    
Gessel also proves a more general result that can be specialized to a sum involving only polynomials of odd degrees to give:
\begin{eqnarray} \label{vinet:eqA.13}    
\fl \sum_{n=0}^{\infty} c_{2n+1} (x;a) \frac{1}{n!} \bigg( -\frac12 at\bigg)^n \nonumber\\
   = \ee^{- at/2}\sum_{\ell=0}^{\infty} \frac{(-x)_{2\ell}}{(1+t)^{2\ell-x}} \bigg[ 1 + \frac{(2\ell - x)}{a(1+t)}\bigg] \frac{1}{\ell !} \bigg( - \frac{y}{2a}\bigg)^{\ell}
     \end{eqnarray}           
    
\subsection{Meixner polynomials~\cite{vinet:ref17}}     \label{vinet:subsectionA.4}  

 The Meixner polynomials $M_n(x;\beta, c)$, $\beta > 0$ and $0 < c < 1$, are also polynomials of a discrete variable. Their 3-term recurrence relation is
\begin{eqnarray} \label{vinet:eqA.14}   
\fl  (c-1)x M_n(x;\beta; c) 
  = c(n+\beta)M_{n+1} (x;\beta;c) \nonumber\\
   - [n + (n+\beta)c] M_n (x;\beta, c) + nM_{n-1}(x;\beta, c).   
    \end{eqnarray}        
The normalized recurrence relation is    
\begin{eqnarray} \label{vinet:eqA.15}   
xp_n(x) = p_{n+1}(x) + \frac{n+(n+\beta)c}{1-c} p_n(x) + \frac{n(n+\beta-1)c}{(1-c)^2} p_{n-1}(x)
    \end{eqnarray} 
where
\begin{eqnarray} \label{vinet:eqA.16}   
M_n(x;\beta, c) = \frac{1}{(\beta)_n} \bigg( \frac{c-1}{c}\bigg)^n p_n(x)
    \end{eqnarray} 
with $(\beta)_n =\beta(\beta + 1) \ldots (\beta + n -1)$.

\subsection{Multivariate Hermite polynomials~\cite{vinet:ref16}}     \label{vinet:subsectionA.5}  

The Hermite polynomials in several variables were introduced by Appell and Kamp\'e de F\'eriet in terms of generating functions. They are denoted by
\begin{eqnarray} \label{vinet:eqA.17}   
H_{\underset{\sim}{n}} (\underset{\sim}{x}) = H_{n_1},\ldots,n_m (x_1, \ldots , x_m)
    \end{eqnarray} 
where $x_i$, $i = 1, \ldots , m$ are the variables and $n_i$ the degrees of the polynomials in the variable $x_i$.

With $A = [a_{ij}]$ a fixed positive definite symmetric square matrix, these polynomials are defined according to
\begin{eqnarray} \label{vinet:eqA.18}   
\exp \left\{ \sum_{i,j=1}^m a_{ij}\bigg(x_i t_j -\frac12 t_it_j\bigg) \right\} = \sum_{n_i=0}^{\infty} \frac{t_1^{n_1}\ldots t_m^{n_m}}{n_1!\ldots n_m!} H_{\underset{\sim}{n}} (\underset{\sim}{x} )
    \end{eqnarray} 
    when $m=1$, we recover the standard Hermite polynomials in one variable for $a_{11} = 2$, We make use of the case $m=2$ in Section~\ref{vinet:section8} of the paper. Note finally that they satisfy a biorthogonality relation with a companion set of polynomials (see \cite{vinet:ref16}).
    
    \section*{References}
    
    \bibliographystyle{unsrt}

\end{document}